\documentclass[10pt,twocolumn,twoside]{IEEEtran}
\IEEEoverridecommandlockouts

\usepackage{cite}
\usepackage{amsmath,amssymb,amsfonts}
\usepackage{algorithmic}
\usepackage{algorithm}
\usepackage{graphicx,booktabs, multirow, array}
\usepackage{textcomp}
\usepackage{balance}
\usepackage{xcolor}
\usepackage[utf8]{inputenc}

\def\BibTeX{{\rm B\kern-.05em{\sc i\kern-.025em b}\kern-.08em
    T\kern-.1667em\lower.7ex\hbox{E}\kern-.125emX}}
\usepackage{lipsum}

\title{ 
Learning-Based Augmentation and Adaptation for Grid Sim-to-Real Model Discrepancy 
}

\author{Sayak Mukherjee, Kyung-Bin Kwon, Ramij R. Hossain, Marcelo Elizondo
\vspace{-0.5cm}
 \thanks{ 
Authors are with Pacific Northwest National Laboratory. The research is supported by the E-COMP (Energy System Co-Design with Multiple Objectives and Power Electronics) Initiative at Pacific Northwest National Laboratory (PNNL). PNNL is operated by Battelle for the U.S. Department of Energy under Contract DEAC05-76RL01830.}
}

\begin{document}
\begingroup
\allowdisplaybreaks

\maketitle

\begin{abstract}
Modern power systems can encounter increased discrepancy between the operators' simulation model and the actual true dynamics of the grid, driven by uncertainties caused by integration of new inverter-based resources (IBRs), large loads, unmodeled dynamics, parameter drifts, etc., to name a few. All of these impact the control room operations, where some critical oscillations may not be captured during the transient studies. To circumvent these issues, we propose a learning-augmented hybrid approach where the operator simulation model is supplemented with artificial intelligence (AI)-learned residual models using the phasor measurement unit (PMU)/ point-on-wave (PoW) based sensed trajectory data. The physics-based operator model provides interpretability and structural consistency, while the learned residual captures discrepancies caused by non-idealities. The learned model employs advanced neural architectures and consists of a backbone encoder and multi-head decoder layers for heterogeneous grid channels. Subsequently, we formulated a continual learning-motivated adaptation framework such that the baseline residual AI model can also be updated when the underlying real grid model changes in future conditions. Extensive numerical simulations are performed on the IEEE 68-bus benchmark model with a diverse set of disturbances, and different state-of-the-art predictive architectures involving recurrent learners, latent neural ODEs, and transformers are explored to demonstrate both residual learning and adaptation capabilities. 
\end{abstract}

\section{Introduction}
Grid operators rely upon the accuracy and precision of their simulation models during dynamic studies. However, integration of different heterogeneous resources such as IBRs and large data center loads has introduced various dynamical uncertainties in the modeling, along with issues related to parameter drifts, proprietary controls, and unmodeled dynamics. System operators rely on previously validated models of grid components and transmission infrastructure, which are maintained through periodic updates. Nevertheless, these models may not fully capture various non-idealities and unmodeled transient dynamics, potentially leading to discrepancies between simulated and actual grid responses during unforeseen critical events. The North American Electric Reliability Corporation (NERC) has reported many critical oscillation events over the years and pointed to some scenarios where the oscillatory responses of the simulation model do not necessarily reproduce the disturbance events \cite{interconnection_oscillation}, and put much emphasis on advanced PMU-based model verifications (\cite{guideline2018power}, NERC MOD 033-3 \cite{mod_33}). Literature on dynamical model validation has recently focused on data-driven approaches in works such as \cite{akhlaghi2020starting, zhang2019generator, huang2013generator}. These observations motivate the development of methods that systematically leverage measured disturbance trajectories to compensate for the mismatches between an operator’s simulation model and the grid’s actual dynamics. In practice, disturbance recordings from phasor measurement units (PMUs) and other monitoring devices are utilized for event playback, allowing the recorded disturbance to be reproduced within the simulation environment and the simulated response to be compared with the measured response. 
However, traditional model calibration can be increasingly difficult when the source of the discrepancy is unknown, when underlying controls are proprietary, or as the system evolves. Simulation-to-real model discrepancy risk can lead to inaccurate predictions, unreliable safety checks, and incorrect stability assessments, especially when unusual events or changes in system conditions occur.

Advances in artificial intelligence (AI) and deep learning have shown that learning-based latent dynamical systems can capture complex nonlinear interactions in physical systems \cite{goodfellow2014generative}. Building on these developments, scientific machine learning integrates domain knowledge with data-driven methods to enhance modeling accuracy and interpretability \cite{cuomo2022scientific, thiyagalingam2022scientific}. Recent works on machine learning-based approaches to accurately model different grid assets include operator-theoretic learning of synchronous generators \cite{moya2023approximating}, physics-informed learning of inverters \cite{KWON2027114177, yang2025data}. On the system level, works such as \cite{hamid2023deep, nandanoori2022graph, stiasny2023physics, moya2023deeponet} focus on constructing deep-learning-based surrogate models of large-scale power-grid dynamics. Purely data-driven models require extensive data and offer limited guarantees in respecting physical laws \cite{karniadakis2021physics}. Physics-guided learning frameworks prioritize incorporating domain knowledge, such as constraints and invariants, as inductive biases to enhance the physical consistency, interpretability, and generalizability of learned models \cite{karpatne2017theory}. Physics-informed learning techniques to power system applications \cite{huang2022applications} include state estimation \cite{iliadis2025physics, su2023probabilistic}, power system dynamic analysis \cite{stiasny2021transient, stiasny2024pinnsim}, power flow calculation \cite{jiang2025unsupervised, donon2020neural}, and optimal power flow \cite{wu2024physics, nellikkath2022physics} etc. However, there has been less focus on directly learning the system-level differences between an operator’s simulation model and actual grid dynamics, particularly when these mismatches change with parameters, grid conditions, and resource characteristics.

This paper proposes a learning-augmented hybrid approach where the operator simulation model is supplemented with artificial intelligence (AI)-learned residual models using the phasor measurement unit (PMU)/ point on wave (PoW) based sensed trajectory data. The physics-based operator baseline model provides interpretability and structural consistency, while the learned residual captures discrepancies caused by non-idealities along with a continual adaptation capability.   The main contributions of this paper are as follows:
\begin{itemize}
   \item We formulate a system-level residual learning framework that retains 
    the existing physics-based operator model and learns the discrepancy between 
    its simulated trajectories and measured grid responses, rather than replacing 
    the complete grid dynamics with a fully data-driven surrogate. We have provided a motivating example with a network of swing oscillators where such model mismatches can lead to contrasting dynamical behaviors. 
    \item We develop a backbone encoder and multi-head decoder-based architecture for heterogeneous grid channels, demonstrated through representative experimentation on angle, 
    frequency, and voltage states.
    \item Subsequently, we formulate a continual learning-motivated adaptation framework such that the baseline residual AI model can also be updated when the underlying real grid model changes. The framework utilizes a pre-trained residual model to be efficiently updated using limited 
    new trajectory data as the underlying grid dynamics evolve, while retaining 
    knowledge of previously learned operating regimes.
    \item We perform extensive numerical studies on the IEEE 68-bus benchmark 
    system under a diverse set of disturbances. Multiple state-of-the-art 
    predictive architectures, including recurrent neural networks, latent neural 
    ODEs, and Transformers, are investigated to demonstrate the effectiveness of 
    the proposed framework for both residual learning and few-shot adaptation.
\end{itemize}

The paper is organized as follows. The nonlinear grid dynamics with simulation-to-real mismatch and a motivating example are presented in Section~II. The detailed learning-based augmentation and adaptation methodologies are described in Section~III. Practical considerations are captured in Section~IV. Section~V shows the numerical experimental details on the simulation model details, data generation schemes, state-of-the-art learners, and extensive AI-based augmentation and adaptation results. Concluding remarks are provided in Section VI.

\section{Nonlinear Grid Dynamics and Operator/Actual Model Mismatch}
We consider the integrated large-scale power grid model using the nonlinear differential-algebraic equations (DAE) described as:
\begin{align}
\dot{x}(t)
&=
f_{\mathrm{act}}
\big(
x(t),y(t),p_{\mathrm{act}}
\big),
\label{eq:actual_dae_diff}
\\
0
&=
g_{\mathrm{act}}
\big(
x(t),y(t),p_{\mathrm{act}}
\big),
\label{eq:actual_dae_alg}
\end{align}
\noindent where $x(t)\in\mathbb{R}^{n_x}$ denotes the dynamic states,
      including generator rotor angles and frequencies,
      inverter states; $y(t)\in\mathbb{R}^{n_y}$ denotes algebraic variables,
      such as bus voltage magnitudes, voltage angles; and $p_{\mathrm{act}}$ denotes the true physical system
      parameters.
In practical settings, the operator model can suffer from discrepancies due to parameter drifts, unmodeled dynamics, topological changes, and inaccurate component and load modeling.
We describe such an operator model with
\begin{align}
\dot{x}(t) &= f_{\mathrm{op}}\big(x(t),y(t), p_{\mathrm{op}}\big),
\label{eq:operator_dae_diff}
\\
0 &= g_{\mathrm{op}}\big(x(t),y(t),p_{\mathrm{op}}\big),
\label{eq:operator_dae_alg}
\end{align}
\noindent where $p_{\mathrm{op}}$ denotes the parameters used in the operator model for computer-aided simulations. The modeling mismatches in both dynamical components, parameters, and algebraic constraints create the deviations between the simulated and real trajectories. In the measurement-based setting, we can consider the solution flow maps induced by the DAE solutions in discrete time, denoted as follows for the actual system:
\begin{equation}
x_{k+1}^{\mathrm{act}}=\mathcal{S}_{\mathrm{act}}(x_k^{\mathrm{act}},p_{\mathrm{act}}).
\label{eq:actual_flow}
\end{equation}
\noindent We consider that the dynamic state estimators (DSEs) capture the relevant state dynamics in the grid. Similarly, the operator's simulation model computes:
\begin{equation}
x_{k+1}^{\mathrm{op}} = \mathcal{S}_{\mathrm{op}}(x_k^{\mathrm{op}}, p_{\mathrm{op}})
\label{eq:operator_flow}
\end{equation}
The solution flow maps $\mathcal{S}_{\mathrm{act}}$ and
$\mathcal{S}_{\mathrm{op}}$ consider the sampled discrete-time version of the actual continuous-time dynamics with the power flow network constraints. Considering such flow map mismatch, instead of learning the full system dynamics, the residual learning paradigm helps in formulating a hybrid model. Let us consider the flow-map residual as:
\begin{equation}
r_{k+1} = \mathcal{S}_{\mathrm{act}}(x_k,p_{\mathrm{act}}) - \mathcal{S}_{\mathrm{op}}(x_k,p_{\mathrm{op}}).
\label{eq:flow_residual}
\end{equation}

\noindent with equivalent residual solution flow operator denoted by $\mathcal{R}(.)$. We consider the critical question:  \textit{how can we compensate for this for accurate prediction and real-time decision-making?} The solution is to learn the approximation residual operator with gathered trajectory data during events:
\begin{equation}
\mathcal{R}_{\theta}: \mathcal{W}_k \rightarrow r_k,
\end{equation}
\noindent where $\mathcal{W}_k$ denotes a finite amount of gathered
states, and can utilize historical data up to the current step, and additional relevant dynamical variables. The corrected predictor becomes
\begin{equation}
\label{eq:hybrid_reconstruction}
\hat{x}_{k+1} = \mathcal{S}_{\mathrm{op}}(\hat{x}_k,p_{\mathrm{op}}) + \mathcal{R}_{\theta}({\mathcal{W}}_k).
\end{equation}
\noindent Therefore, the learning paradigm utilizes the baseline physics models and tries to optimize for the discrepancies arising from unmodeled dynamics, parameter deviations, structural changes, inverter-control inconsistencies, and other nonlinear phenomena. As an example, a real grid with lower damping than the operator-calibrated model might exhibit oscillations that the operator model does not predict, making it necessary for the residual learner to infer such discrepancies. Moreover, we investigate continual adaptation of the learned residual augmentation using limited data as grid dynamics evolve over time. The following case study illustrates this problem.
\subsection*{A Motivating Example:}
\begin{figure}[t!]
    \centering
    \includegraphics[width=0.98\linewidth, trim=0 15pt 0 0, clip]{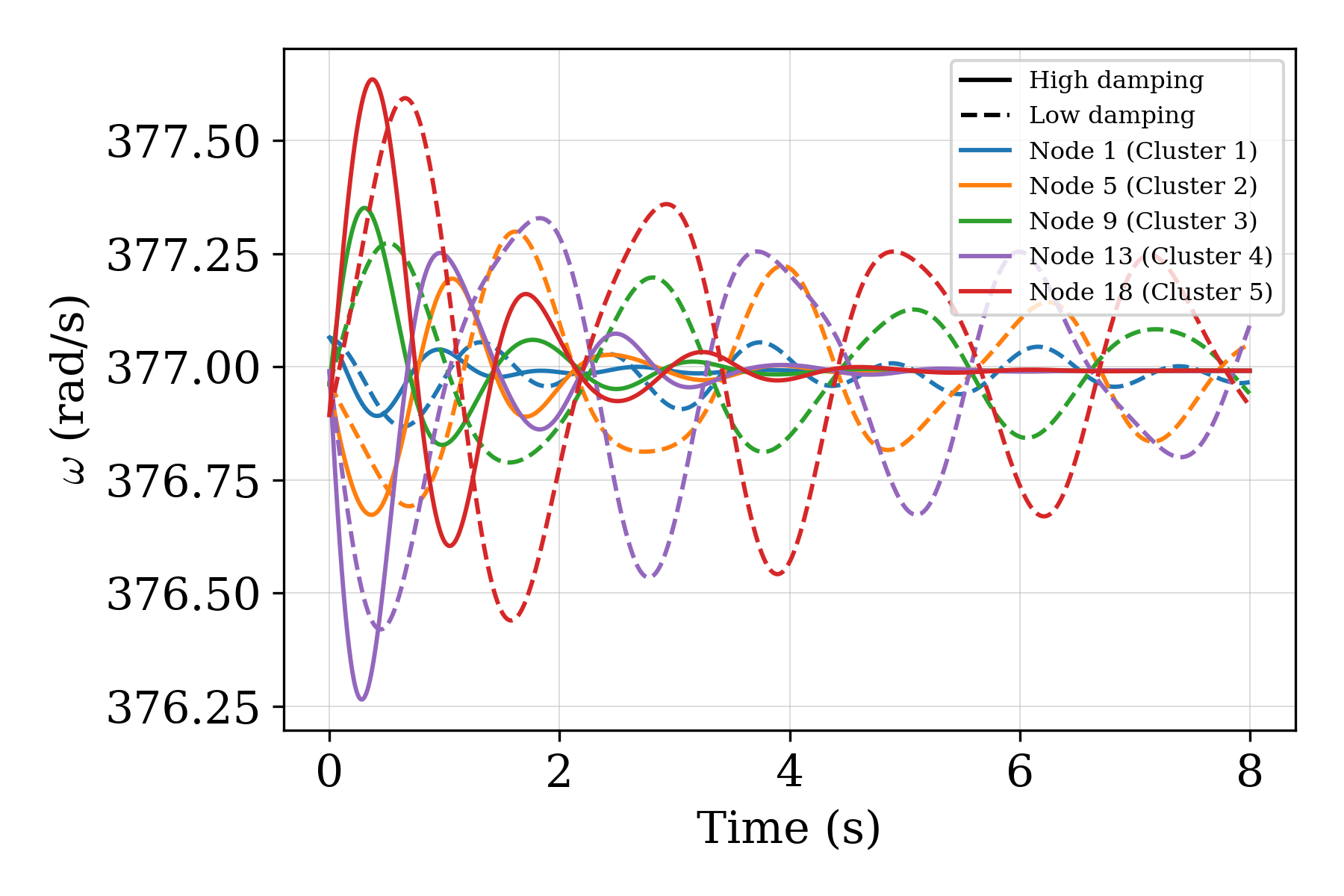}
    \caption{Model differences between the actual model and the simulator model lead to significant performance discrepancies}
    \label{fig:motivating_differences}
\end{figure}
\begin{figure}[t!]
    \centering
    \includegraphics[width=0.98\linewidth, trim=0 15pt 0 0, clip]{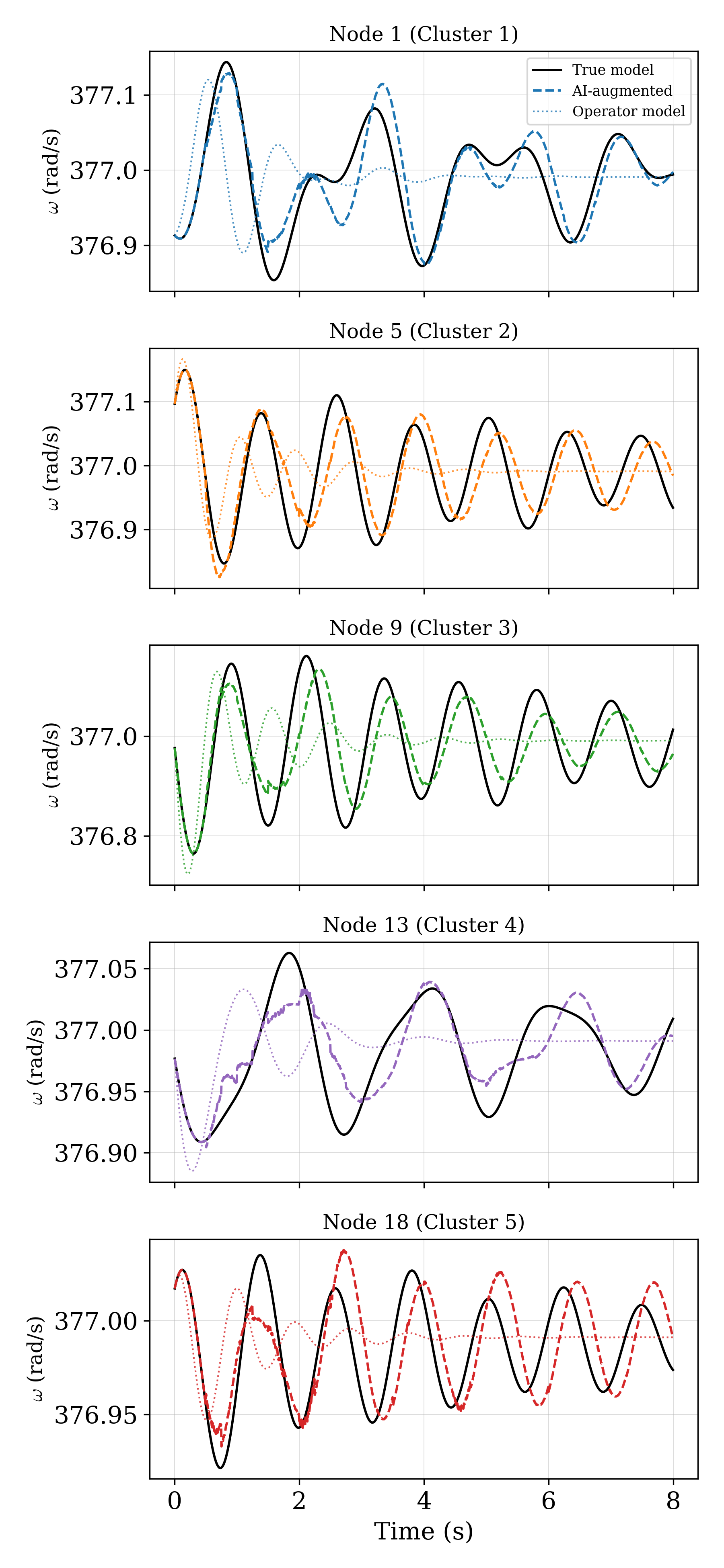}
    \caption {AI augmentation helps recover the actual dynamics from the operator's simulation model}
    \label{fig:motivating_performances}
    \vspace{-0.5 cm}
\end{figure}
We consider a synthetic network of interconnected swing oscillators that mimic the synchronous generator dynamics,  consisting of $N = 20$ synchronous machines
organized into $C = 5$ clusters of 4 machines each.
The dynamics of each machine $i$ are governed by the swing equation.
\begin{equation}
    M_i\,\ddot{\delta}_i \;=\; P_{m,i} - P_{l,i} - D_i\,\dot{\delta}_i
    - \sum_{j \neq i} K_{ij}\sin(\delta_i - \delta_j),
\end{equation}
where $\delta_i$ is the rotor angle, $\dot{\delta}_i = \omega_i$ the frequency deviation,
$M_i$ the inertia constant, $D_i$ the damping coefficient,
$P_{m,i}$ the mechanical power input, $P_{l,i}$ the electrical load,
and $K_{ij}$ the coupling strength between machines $i$ and $j$.
The network has a two-level coupling structure: intra-cluster links are
stronger than inter-cluster links.
Mechanical powers and loads are drawn randomly with the global power balance
$\sum_i (P_{m,i} - P_{l,i}) = 0$ enforced.

Two operating regimes are studied to contrast damping behavior.
In the \emph{high-damping} scenario (System~1), the damping coefficient is set to $D = 2.0$ with intra- and inter-cluster coupling strengths
$K_{\mathrm{intra}} = 8.0$ and $K_{\mathrm{inter}} = 1.0$, respectively.
Under these conditions, the rotor frequencies return to the nominal value
$\omega_0 = 2\pi \times 60~\text{rad/s}$ within approximately $3~\text{s}$.
In the \emph{low-damping} scenario (System~2), the damping is reduced to
$D = 0.25$ and the coupling strengths are weakened to
$K_{\mathrm{intra}} = 5.0$ and $K_{\mathrm{inter}} = 0.4$,
resulting in sustained inter-area oscillations that persist well beyond $8~\text{s}$, showing the significant discrepancies as in Fig. \ref{fig:motivating_differences}.
All machines have $M_i = 1$.
Initial rotor angles are set with cluster-level offsets
$\boldsymbol{\delta}^{(k)}_0 \in \{0.0,\,0.1,\,-0.1,\,0.2,\,-0.2\}~\text{rad}$
plus small random perturbations $\mathcal{N}(0,\,0.02^2)$,
and initial frequency deviations are drawn from $\mathcal{N}(0,\,0.05^2)$.
Trajectories are simulated using a stiff ODE integrator
with tolerances $r_{\mathrm{tol}} = 10^{-6}$ and $a_{\mathrm{tol}} = 10^{-8}$. Subsequently, the AI-augmented model follows a \emph{residual correction} framework.
We will describe the details of the methodology in the subsequent section.
Specifically, we train a direct multi-step LSTM that encodes a history and predicts multiple steps of residual states in a single forward pass,
avoiding the error accumulation of recursive one-step rollouts.
At inference, the corrected state is recovered by supplementing the operator model outputs with the AI-residual prediction, which shows much enhanced performance as in Fig.~\ref{fig:motivating_performances}.
This hybrid formulation retains the interpretability and physical structure of the
operator model while using the neural network solely to correct its systematic errors.

\section{Learning and Adaptation Methodology}

The residual learner can be trained by minimizing the discrepancy between the corrected rollout and the actual trajectory. A one-step residual loss is
\begin{equation}
    \mathcal{L}_{\mathrm{1step}}(\theta)
    \!=\!
    \frac{1}{N}\!
    \sum_{i=1}^{N}\!
    \sum_{k}\!
    \left\|
    x_{k+1}^{(i),\mathrm{act}}
    \!-\!
    \mathcal{S}_{\mathrm{op}}\big(x_k^{(i)},p_{\mathrm{op}}\big)
    \!-\!
    \mathcal{R}_{\theta}\big(\mathcal{H}_k^{(i)}\big)
    \right\|_2^2 .
\end{equation}
However, for dynamic stability assessment and long-horizon prediction, accurate one-step prediction is not sufficient. Therefore, we also consider autoregressive rollout:
\begin{align}
    \hat{x}_{k+1}
    &=
    \mathcal{S}_{\mathrm{op}}(\hat{x}_k,{p_{\mathrm{op}}})
    +
    \mathcal{R}_{\theta}(\hat{\mathcal{H}}_k), \\
& \hat{\mathcal{H}}_k
=
\left\{
\hat{r}_{k-L+1},
\hat{r}_{k-L+2},
\ldots,
\hat{r}_{k}
\right\}.
\end{align}
The corresponding rollout loss over a horizon $T_r$ is
\begin{equation}
    \mathcal{L}_{\mathrm{rollout}}(\theta)
    =
    \frac{1}{N}
    \sum_{i=1}^{N}
    \sum_{k=1}^{T_r}
    \left\|
    x_k^{(i),\mathrm{act}}
    -
    \hat{x}_k^{(i)}
    \right\|_2^2 .
\end{equation}

\subsection{Multi-head Architecture for Heterogeneous Grid Channels}

In grid dynamical models, we can have states that are characteristically distinct, such as generator angles, frequencies, and then inverter internal voltage states; then the residual vector is decomposed as
\begin{equation}
{r}_k =
\begin{bmatrix}
{r}^{(a)}_k \\
{r}^{(f)}_k \\
{r}^{(v)}_k
\end{bmatrix},
\label{eq:state_partition}
\end{equation}
Here, \({r}^{(a)}_k\), \({r}^{(f)}_k\), and \({r}^{(v)}_k\) denote the angle, frequency, and voltage residual vectors, respectively. We consider this as our working example for this paper; however, this idea of characteristically distinct state groups can be extended to a larger set of state variable classes. 
Each trajectory is sampled over \(T\) time instants on the interval \([0, L]\) s. For each trajectory, the residual sequence is
$
\{{r}_0,{r}_1,\ldots,{r}_{T-1}\}, 
\widetilde{{r}}_k = \alpha {r}_k.
\label{eq:global_scaling}
$

The training task is formulated as next-step sequence prediction. Specifically, the input and target sequences used for training are:
\begin{equation}
{X} =
\left[
\widetilde{{r}}_0,
\widetilde{{r}}_1,
\ldots,
\widetilde{{r}}_{T-2}
\right],
\label{eq:X_def}
\end{equation}
\begin{equation}
{Y} =
\left[
\widetilde{{r}}_1,
\widetilde{{r}}_2,
\ldots,
\widetilde{{r}}_{T-1}
\right].
\label{eq:Y_def}
\end{equation}

Because the residual channels can differ significantly in magnitude across state groups, per-state normalization is applied using statistics computed only from the training set. Let \(\mu_j\) and \(\sigma_j\) denote the mean and standard deviation of the \(j\)-th scaled residual component over all training trajectories and time indices. Then the normalized residual component is
$
\widehat{r}_{k,j} =
(\widetilde{r}_{k,j} - \mu_j)/\sigma_j,
\; j = 1,\ldots,n_x.
\label{eq:normalization}
$

Applying such normalization to every state dimension yields the normalized input-target pair.
$
\widehat{{X}},\widehat{{Y}} \in \mathbb{R}^{N \times (T-1) \times n_x},
$
where \(N\) is the number of trajectories. 
To model the temporal evolution of the residual sequence, a shared-backbone multi-head learner ( e.g., long short-term memory (LSTM)) network is employed. Given an input sequence \(\widehat{{X}}\), the learner's encoder layer produces a hidden/latent sequence
\begin{equation}
{H} = \mathcal{L}(\widehat{{X}}),
\label{eq:lstm_hidden}
\end{equation}
where
${H} \in \mathbb{R}^{N \times (T-1) \times h}$, where $h$ is the dimension of the hidden state representation.  For example, in one of our numerical implementations, the learner's backbone consists of \(5\) LSTM layers with a dropout rate of \(0.1\). The shared hidden representation \({H}\) is then passed to the channel-specific output heads:
\begin{align}
\widehat{{Y}}^{(a)} = \mathcal{G}_a({H}),
\widehat{{Y}}^{(f)} = \mathcal{G}_f({H}), 
\widehat{{Y}}^{(v)} = \mathcal{G}_v({H}). 
\end{align}

In our baseline implementation, the example angle and voltage heads are composed of linear projections:
\begin{equation}
\mathcal{G}_a({H}) = {W}_a {H} + {b}_a,
\qquad
\mathcal{G}_v({H}) = {W}_v {H} + {b}_v.
\label{eq:linear_heads}
\end{equation}

To provide additional representational capacity for the frequency states, a deeper nonlinear head is used:
\begin{equation}
\mathcal{G}_f({H}) =
{W}_{f,2}\,
\phi\!\left({W}_{f,1}{H} + {b}_{f,1}\right)
+ {b}_{f,2},
\label{eq:deep_freq_head}
\end{equation}
where \(\phi(\cdot)\) is the ReLU activation function. The final normalized residual prediction is obtained by concatenating the three group predictions in the original state ordering:
\begin{equation}
\widehat{{Y}}^{\mathrm{pred}} =
\left[
\widehat{{Y}}^{(a)},
\widehat{{Y}}^{(f)},
\widehat{{Y}}^{(v)}
\right]
\in
\mathbb{R}^{N \times (T-1) \times n_x}.
\label{eq:concat_output}
\end{equation}

To account for the heterogeneous complexity of different state groups, the training objective is defined as a weighted sum of group-wise mean squared errors (MSEs). Let
$
\widehat{{Y}}^{(a)}_{\mathrm{act}}, \quad
\widehat{{Y}}^{(f)}_{\mathrm{act}}, \quad
\widehat{{Y}}^{(v)}_{\mathrm{act}}
$
denote the normalized target sequences for the three groups. The corresponding losses are
\begin{align}
\mathcal{L}_a &=
\mathrm{MSE}\!\left(
\widehat{{Y}}^{(a)},
\widehat{{Y}}^{(a)}_{\mathrm{act}}
\right), \label{eq:La}\\
\mathcal{L}_f &=
\mathrm{MSE}\!\left(
\widehat{{Y}}^{(f)},
\widehat{{Y}}^{(f)}_{\mathrm{act}}
\right), \label{eq:Lf}\\
\mathcal{L}_v &=
\mathrm{MSE}\!\left(
\widehat{{Y}}^{(v)},
\widehat{{Y}}^{(v)}_{\mathrm{act}}
\right). \label{eq:Lv}
\end{align}
The overall training loss is
\begin{equation}
\mathcal{L} =
\lambda_a \mathcal{L}_a +
\lambda_f \mathcal{L}_f +
\lambda_v \mathcal{L}_v,
\label{eq:total_loss}
\end{equation}
with corresponding scaling factors.

The predicted residual sequence is combined with the operator trajectory using \eqref{eq:hybrid_reconstruction}. To ensure consistency at the initial time step, the true residual at \(k=0\) is retained:
\begin{equation}
{r}^{\mathrm{pred}}_{\mathrm{full}} =
\left[
{r}_0,\;
{r}^{\mathrm{pred}}_1,\;
{r}^{\mathrm{pred}}_2,\;
\ldots,\;
{r}^{\mathrm{pred}}_{T-1}
\right].
\label{eq:initial_padding}
\end{equation}
The resulting hybrid trajectory is therefore computed as:
\begin{equation}
{x}^{\mathrm{hyb}}_k =
{x}^{\mathrm{op}}_k + {r}^{\mathrm{pred}}_{\mathrm{full},k},
\qquad k=0,\ldots,T-1.
\label{eq:final_hybrid}
\end{equation}

\subsection{Continual Adaptation of Residual Dynamics Models}

Next, we consider the scenario where the actual grid model evolves, and therefore, the AI-based residual model needs to be adapted. For ease of exposition, we denote the solution flow map generically as:
\begin{equation}
    x_{k+1}^{(i)} = \mathcal{S}^{(i)}(x_k^{(i)}), \quad i \in \{1,2,3\},
\end{equation}
where:
\begin{itemize}
    \item System 1: $\mathcal{S}_{\mathrm{op}}$, operator model (physics-based, imperfect),
    \item System 2: $\mathcal{S}_{\mathrm{act}}$, base ``actual'' system (used for pre-training),
    \item System 3: $\mathcal{S}_{\mathrm{new}}$, new system with limited data (few-shot adaptation target).
\end{itemize}

For the adaptation scenario, we consider that the trajectories stored for the modified system map $\mathcal{S}_{\mathrm{new}}$ are of limited amount. Motivating research works include various fundamental principles on adapting selected layers \cite{yosinski2014transferable}, parameter-efficient transfer learning \cite{houlsby2019parameter}, reducing catastrophic forgetting \cite{kirkpatrick2017overcoming}, and continual learning \cite{rolnick2019experience}. We describe the steps utilized for adapting the residual augmentation modules subsequently.    

\subsubsection{Pre-training (On System 2)}

A neural baseline model $\mathcal{R}_\theta$ is trained on the System~2 residuals as described in the previous subsection:
\begin{equation}
    \mathcal{R}_\theta : \mathbb{R}^{h \times d} \rightarrow \mathbb{R}^{p \times d},
\end{equation}
with training loss:
\begin{equation}
    \mathcal{L}_{\text{pre-train}} =
    \mathbb{E}
    \left[
    \left\|
    \mathcal{R}_\theta(r^{(2)}_{k-h:k-1}) - r^{(2)}_{k:k+p-1}
    \right\|^2
    \right].
\end{equation}

The model consists of: 1) an encoder (learner backbone), 2) a prediction head mapping latent features to future residuals. Individual model details of the SOTA learners will be provided shortly in the experiment section.

\subsubsection{Few-Shot Adaptation (Based on the New System 3)}

Given limited System 3-generated event data, we adapt the pre-trained model.
First, we compute the residuals $r_k^{(3)}$ using the limited System~3 event data and the baseline operator simulation-based data ($\tilde{x}_k$) : 
\begin{equation}
    r_k^{(3)} = x_k^{(3)} - \tilde{x}_k.
\end{equation}
The adaptation objective is then built using both new data and some amount of pre-training data.
We fine-tune $\mathcal{R}_\theta$ using:
\begin{equation}
    \mathcal{L}_{\text{adapt}} =
    \mathcal{L}_{\text{new}} +
    \lambda_{\text{replay}} \mathcal{L}_{\text{replay}},
\end{equation}
where:
\begin{align}
    \mathcal{L}_{\text{new}} &=
    \mathbb{E}
    \left[
    \left\|
    \mathcal{R}_\theta(r^{(3)}_{k-h:k-1}) - r^{(3)}_{k:k+p-1}
    \right\|^2
    \right], \\
    \mathcal{L}_{\text{replay}} &=
    \mathbb{E}
    \left[
    \left\|
    \mathcal{R}_\theta(r^{(2)}_{k-h:k-1}) - r^{(2)}_{k:k+p-1}
    \right\|^2
    \right].
\end{align}

The replay term mitigates catastrophic forgetting of System 2 dynamics.
To ensure stable adaptation under limited data, we freeze the encoder parameters (can also include an additional initial set of layers when needed), i.e.,
$
    \theta = (\theta_{\text{enc}}, \theta_{\text{head}}),
$ $\theta_{\text{enc}}$ is considered to be fixed, and $\theta_{\text{head}}$ are optimized.  
Thus, adaptation modifies only the output mapping:
\begin{equation}
    z = \text{Encoder}(r_{k-h:k-1}; \theta_{\text{enc}}),
\end{equation}
\begin{equation}
    \hat{r}_{k:k+p-1} = \text{Head}(z; \theta_{\text{head}}).
\end{equation}

At inference, we perform chunked autoregressive rollout. Given initial residual history:
$
    \hat{r}_{0:h-1} = r_{0:h-1},
$
we iteratively apply:
\begin{align}
    \hat{r}_{t:t+p-1} = \mathcal{R}_\theta(\hat{r}_{t-h:t-1}), 
    t \leftarrow t + p,
\end{align}
until the full trajectory is reconstructed.
The final predicted state is:
$
    \hat{x}_k = \tilde{x}_k + \hat{r}_k.
$
Thus, the learned model acts as a correction to the operator dynamics.

\begin{algorithm}[h!]
\caption{Learning-Based Augmentation and Few-Shot Adaptation}
\label{alg:residual_adaptation}
\begin{algorithmic}[1]
\STATE \textbf{Input:} Operator simulator model $\mathcal{S}_{\mathrm{op}}$, measured actual trajectories
$\mathcal{D}_{2}$, limited adaptation trajectories $\mathcal{D}_{3}$,
history length $h$, prediction horizon $p$, replay weight $\lambda_{\mathrm{replay}}$
\STATE \textbf{Output:} Adapted hybrid predictor $\hat{\mathcal{S}}_{\mathrm{hyb}}$

\STATE \textbf{Pre-training with baseline residual:}
\FOR{each disturbance trajectory in $\mathcal{D}_{2}$}
    \STATE \textit{Simulate} the operator model with the disturbance replay under the same operating condition.
    \STATE \textit{Compute} residual trajectory
    $
    r_k^{(2)} = x_k^{(2)} - x_k^{\mathrm{op}}.
    $
\ENDFOR

\STATE \textit{Normalize} residual channels, and  \textit{group} residual states into physically distinct sets
(e.g., angle, frequency, and voltage channels).

\STATE \textit{Construct} history--future training pairs
$
r_{k-h:k-1}^{(2)}
\rightarrow
r_{k:k+p-1}^{(2)}.
$
\STATE \textit{Train} shared encoder parameters $\theta_{\mathrm{enc}}$ and
group-specific prediction head parameters $\theta_{\mathrm{head}}$ by minimizing
$
\mathcal{L}_{\mathrm{pre}}
=
\sum_{j \in \{{a,f,v}\}}
\lambda_j \mathcal{L}_j.
$

\STATE \textit{Construct} the augmented predictor
$
\hat{x}_k
=
x_k^{\mathrm{op}} + \hat{r}_k.
$

\STATE \textbf{Few-shot adaptation when needed:}
\IF{new disturbance trajectories $\mathcal{D}_{3}$ become available}
    \STATE \textit{Generate} the corresponding operator-model trajectories.
    \STATE \textit{Compute} new residuals
    $
    r_k^{(3)} = x_k^{(3)} - x_k^{\mathrm{op}}.
    $
    \STATE \textit{Freeze} selected initial-layer/encoder parameters.
    \STATE \textit{Update} only the selected later-layers/ adapters/prediction heads using
    \[
    \mathcal{L}_{\mathrm{adapt}}
    =
    \mathcal{L}_{\mathrm{new}}
    +
    \lambda_{\mathrm{replay}}
    \mathcal{L}_{\mathrm{replay}}.
    \]
\ENDIF

\STATE \textbf{Inference:}
\STATE \textit{Predict} recursively future residuals given an initial residual history, 
$
\hat{r}_{t:t+p-1}
=
\mathcal{R}_{\theta}\!\left(
\hat{r}_{t-h:t-1}
\right).
$
\STATE \textit{Reconstruct} the corrected grid trajectory as
$
\hat{x}_k
=
x_k^{\mathrm{op}} + \hat{r}_k.
$

\RETURN $\hat{\mathcal{S}}_{\mathrm{hyb}}$
\end{algorithmic}
\end{algorithm}

Alg. \ref{alg:residual_adaptation} summarizes the steps of the overall methodology. This approach leverages a physics-informed structure, where the operator establishes the baseline dynamics, ensuring that the model is grounded in fundamental principles. Data efficiency is achieved by focusing learning solely on the residuals, which minimizes the amount of data required and streamlines the training process. Few-shot adaptability is enabled by updating only the head parameters, allowing the model to quickly adjust to new scenarios with minimal additional information. Prediction accuracy is maintained through direct multi-step prediction, which helps reduce compounding errors that often arise in sequential forecasting. Finally, continual learning is supported by replay mechanisms that preserve previously learned regimes, ensuring that the model retains valuable knowledge as it encounters new data and environments.

\section{Practical Considerations}

\begin{itemize}

\item \textit{Availability of the baseline operator model:}
The proposed framework does not require replacing the existing physics-based grid simulator. Instead, it assumes that the operator model can be simulated for the same operating conditions and for the associated event as captured in the measurement data. This is practically attractive since transmission operators already maintain dynamic simulation models for planning and stability studies. The learned model therefore serves as an augmentation layer that compensates for systematic discrepancies in the existing model.

\item \textit{Collection of disturbance trajectory data:}
The AI-augmentation and adaptation process requires time-synchronized trajectories representing the actual system response during sufficiently informative disturbances. Such data can potentially be obtained from PMUs, digital fault recorders, or high-resolution point-on-wave measurement (PoW) units. Along with the physical disturbance data logs, depending on the feasibility, the operator can augment them with additional commissioning-controlled experimentation. Furthermore, to tackle the issue of a limited amount of datasets encompassing sufficient disturbances, our approach focuses only on the discrepancy relative to an available physics-based model rather than re-learning the complete system dynamics during pre-training. Once the initial sufficiently excited datasets are utilized for the baseline pre-training, the subsequent few-shot adaptation mechanism is built by utilizing a limited amount of new transient data when the underlying system changes.

\item \textit{Alignment between measurements and simulations:}
For constructing the residual component, the operator simulation-based and measured disturbance data should approximately correspond to the same initial power flow, disturbance, and other grid configurations of topology and control settings. The implementation requires employing highly accurate event characterization, playback, and operating point reconstruction to minimize the occurrence of incorrect residuals and model discrepancy. Such event alignment-based pre-processing therefore constitutes an important practical step before training.

\item \textit{Measurement noise and missing channels:}
Field measurements are often affected by sensor noise, communication errors, missing samples, and insufficient access to internal device states. To ensure reliable results, practical implementations should include robust data-quality screening, synchronization, filtering, and strategies for handling missing measurements. When internal states cannot be accessed, the residual learner can be designed to use observable quantities, as long as the chosen measurements provide sufficient information about the relevant system dynamics.

\item \textit{Changes in operating conditions and system configuration:}
In real-world scenarios, the modeling mismatch between the operator and actual systems may vary with changes in variable generation and loads, interconnection topology, changes in controller settings, and equipment changes. Consequently, a residual model trained under one operating regime may not remain accurate for all time. Online performance monitoring and periodic adaptation using newly collected events are therefore essential for maintaining model fidelity.

\item \textit{Safe adaptation and model validation:}
Although the proposed framework enables continual update of the learnable residual layers, this may not lead to deployment of the updated model for operational studies. Instead, newly adapted models must first be evaluated against held-out disturbance events and compared to the previous nominal operator model. Before incorporating an updated residual model into the simulation workflow, appropriate acceptance criteria, such as thresholds for trajectory errors, can be applied to ensure reliability and performance.

\item \textit{Computational and deployment feasibility:}
Training and adaptation can be performed offline following the collection of disturbance data, while inference only requires evaluation of the learned residual model together with the existing operator simulation. Thus, the proposed framework does not necessarily require real-time neural-network training or modification of the underlying simulation engine. Depending on the intended application, the learned correction can be integrated as a post-processing predictor or more tightly coupled with the simulation workflow.
\end{itemize}

\section{Numerical Experiments}

\subsection{Inverter-integrated IEEE Benchmark 68-bus Model and Data Generation}

We study a bulk power system comprising $N$ buses, with $m$ synchronous generators (SGs) and $n$ grid-forming (GFM) inverters. Each SG's dynamics are captured by the standard classical swing equation model \cite{kundur}:
\begin{align}
&\dot{\delta}_i = \omega_i - \omega_0,\\
&\dot{\omega}_i = \tfrac{1}{M_i}\left[D_i(\omega_0 - \omega_i) + P_i - P_{ei}\right],
\label{eq:sg}
\end{align}
where $\delta_i$ and $\omega_i$ are, respectively, the rotor angle and rotor frequency of generator $i$; $P_i$ denotes the mechanical input power; and $P_{ei}$ is the electrical output power. The inertia and damping coefficients are given by $M_i$ and $D_i$, respectively.

The GFM inverters are represented using P-f and Q-V droop control, following the WECC-approved positive-sequence $REGFM\_A1$ model \cite{REGFM_A1} originally developed at PNNL. Their dynamics are given by \cite{REGFM_A1}:
\begin{subequations}
\label{eq:gfm}
\begin{align}
&\dot{\delta}_j = \omega_j - \omega_0,\\
&\dot{\omega}_j = \frac{1}{\tau_j}\left[\omega_0 - \omega_j + m_{p_j}(P_j^{set} - P_j)\right], \label{omega_eq}\\
&\dot{V}^e_j = \frac{1}{\tau_j}\left[V_j^{set} - V_j - V^e_j + m_{q_j}(Q_j^{set} - Q_j)\right],\\
&\dot{E}_j = k^{pv}_j \dot{V}^e_j + k^{iv}_j V^e_j,
\end{align}
\end{subequations}
where $\delta_j$, $\omega_j$, $V_j$, and $E_j$ denote the voltage angle, frequency, terminal voltage magnitude, and internal voltage magnitude of GFM $j$, respectively. $P_j^{set}$, $Q_j^{set}$, and $V_j^{set}$ are the active-power, reactive-power, and voltage setpoints, while $P_j$ and $Q_j$ are the corresponding power injections measured at the GFM's point of connection. The auxiliary variable $V^e_j$ represents the voltage error signal. The droop gains $m_{p_j}$ and $m_{q_j}$ set the $P-f$ droop and $Q-V$ droop response, respectively, and $\tau_j$ is the measurement filter time constant (set to $0.01$ s in this study). The constants $k^{pv}_j$ and $k^{iv}_j$ are the proportional and integral gains of the $Q-V$ droop controller.

The active and reactive power balance at each bus $j$, for $j = 1,\dots,m$, is given by:
\begin{subequations}
\label{eqn:load_flows}
\begin{align}
\label{eqn:load_active}
0 &= P_{ej} - {\rm Re}\left\{ \sum\limits_{k = 1,\,k \ne j}^N V_j \left( V_{jk} B_{jk} \right)^* \right\} - V_j^2 G_j,\\
\label{eqn:load_reactive4}
0 &= Q_{ej} - {\rm Im}\left\{ \sum\limits_{k = 1,\,k \ne j}^N V_j \left( V_{jk} B_{jk} \right)^* \right\} - V_j^2 B_j,
\end{align}
\end{subequations}
where $G_j$ and $B_j$ are the shunt conductance and susceptance at bus $j$ (accounting for line charging), and $B_{jk}$ is the susceptance of the (lossless) tie-line connecting buses $j$ and $k$. These power-flow relations are summarized compactly as $0 = g(x_s, x_f, V)$.

To evaluate the proposed learning and adaptation methodology, we synthetically construct the generic operator model and the actual physical system by introducing deliberate parameter discrepancies. Specifically, the mismatch between the two systems is established by varying critical dynamic parameters: the inverter droop coefficients ($m_{p_j}$, $m_{q_j}$), synchronous generator inertia constants ($M_i$), and damping coefficients ($D_i$). For instance, the operator model assumes nominal baseline parameters (e.g., 1\% droop, 0.15 damping coefficient, and 1.2 p.u. inertia), whereas the actual system exhibits a more poorly damped, low-inertia behavior (e.g., 2.1\% droop, 0.05 damping coefficient, and 0.7 p.u. inertia).

To systematically generate a diverse set of dynamic trajectories for training and testing, we conduct comprehensive load perturbation experiments. Sudden load step changes with randomly sampled magnitudes are applied at various randomly selected bus locations across the test system. For each perturbation scenario, time-domain simulations are executed for both the operator and actual models to capture the transient evolution of bus frequencies and relative voltage angles. The collected multi-channel trajectories are then utilized to compute the residual sequences. These sequences are defined as the difference between the actual and operator responses and serve as the training targets for our predictive learner. Furthermore, to validate the few-shot adaptability of the proposed framework, a supplementary adaptation dataset is generated. This dataset simulates a subsequent parameter drift in the actual system with a very limited number of trajectory samples, thereby testing the model's capability to rapidly adjust to newly emerging dynamic conditions.
Specifically, we generated 1,000 trajectory samples for the pre-training phase, and a restricted set of only $50$ trajectory samples for the adaptation phase. Each trajectory consists of 1,000 time steps sampled at a resolution of $0.01$~s (yielding a $10$-second simulation window).

\subsection{State-of-the-art Predictive Learners}
We have tested different state-of-the-art predictive algorithms to learn the residual solution flow operator $\mathcal{R}_{\theta}(.)$. 

\subsubsection{LSTM-Based Residual Predictive Learner}
Rather than directly learning the full system dynamics, the predictive
learner approximates the temporal evolution of the residual sequence.

\subsubsection*{LSTM Encoder}

For each residual vector $r_k$ in the history window, the LSTM updates
its hidden state $h_k$ and cell state $c_k$. The LSTM equations are
\begin{align}
    i_k &= \sigma(W_i r_k + U_i h_{k-1}+b_i),\\
    f_k &= \sigma(W_f r_k + U_f h_{k-1}+b_f),\\
    o_k &= \sigma(W_o r_k + U_o h_{k-1}+b_o),\\
    \tilde c_k &= \tanh(W_c r_k + U_c h_{k-1}+b_c),\\
    c_k &= f_k\odot c_{k-1}+i_k\odot\tilde c_k,\\
    h_k &= o_k\odot\tanh(c_k),
\end{align}
where $i_k$, $f_k$, and $o_k$ denote the input, forget, and output
gates, respectively, $\sigma(\cdot)$ is the sigmoid activation
function, and $\odot$ denotes element-wise multiplication.

After processing the complete history window, the final hidden
representation 
\begin{equation}
    z_t = h_t
\end{equation}
provides a latent representation of the recent residual dynamics.

\subsubsection*{Multi-Head Direct Multi-Step Prediction}

The residual vector is partitioned according to the physical state
groups as described before,
\begin{equation}
    r_t =
    \begin{bmatrix}
        r_t^{a}\\
        r_t^{f}\\
        r_t^{v}
    \end{bmatrix},
\end{equation}
where $r_t^{a}$, $r_t^{f}$, and $r_t^{v}$ correspond to
angle, frequency, and voltage-related residual states, respectively. A shared LSTM representation $z_t$ is passed to three specialized
prediction heads,
\begin{align}
    \widehat{\mathcal{R}}_{t+1:t+P}^{a}
        &= \phi_{a}(z_t),\\
    \widehat{\mathcal{R}}_{t+1:t+P}^{f}
        &= \phi_{f}(z_t),\\
    \widehat{\mathcal{R}}_{t+1:t+P}^{v}
        &= \phi_{v}(z_t),
\end{align}
where $\phi_{a}$, $\phi_{f}$, and $\phi_v$ are feedforward
neural-network prediction heads. Importantly, each head predicts the entire $P$-step future horizon
directly rather than recursively generating individual time steps.
The complete predicted residual chunk is reconstructed as
\begin{equation}
    \widehat{\mathcal{R}}_{t+1:t+P}
    =
    \left[
    \widehat{\mathcal{R}}_{t+1:t+P}^{a},
    \widehat{\mathcal{R}}_{t+1:t+P}^{f},
    \widehat{\mathcal{R}}_{t+1:t+P}^{v}
    \right].
\end{equation}

\subsubsection{Neural ODE-Based Residual Predictor}

The residual history is first mapped to a latent initial condition
using a recurrent encoder,
\begin{equation}
    z_0
    =
    \mathcal{E}_{\phi}
    \left(
    \mathcal{W}_t
    \right),
    \qquad
    z_0\in\mathbb{R}^{n_z},
\end{equation}
where $\mathcal{E}_{\phi}$ is a GRU-based encoder. The temporal evolution of the latent representation is then modeled
using a Neural ODE,
\begin{equation}
    \frac{d z(\tau)}{d\tau}
    =
    f_{\theta}\big(z(\tau)\big),
    \qquad
    z(0)=z_0,
\end{equation}
where $f_{\theta}$ is a neural network representing the latent vector
field. Equivalently, the evolved latent state is
\begin{equation}
    z(\tau_f)
    =
    z_0+
    \int_{0}^{\tau_f}
    f_{\theta}\big(z(\tau)\big)\,d\tau.
\end{equation}

The resulting latent state is passed through separate prediction heads
for the angle, frequency, and voltage-related residuals,
\begin{align}
    \widehat{\mathcal{R}}^{a}_{t+1:t+P}
        &= \mathcal{D}_{a}\big(z(\tau_f)\big),\\
    \widehat{\mathcal{R}}^{f}_{t+1:t+P}
        &= \mathcal{D}_{f}\big(z(\tau_f)\big),\\
    \widehat{\mathcal{R}}^{v}_{t+1:t+P}
        &= \mathcal{D}_{v}\big(z(\tau_f)\big).
\end{align}

The complete future residual prediction is therefore
\begin{equation}
    \widehat{\mathcal{R}}_{t+1:t+P}
    =
    \operatorname{concat}
    \left(
    \widehat{\mathcal{R}}^{a}_{t+1:t+P},
    \widehat{\mathcal{R}}^{f}_{t+1:t+P},
    \widehat{\mathcal{R}}^{v}_{t+1:t+P}
    \right).
\end{equation}

The model parameters are learned by minimizing the weighted
multi-step prediction error as described before,
\begin{equation}
    \mathcal{L}
    =
    \lambda_{a}\mathcal{L}_{a}
    +
    \lambda_{f}\mathcal{L}_{f}
    +
    \lambda_v\mathcal{L}_{v},
\end{equation}
where
\begin{equation}
    \mathcal{L}_{j}
    =
    \frac{1}{P n_j}
    \sum_{k=1}^{P}
    \left\|
    \hat r_{t+k}^{j}-r_{t+k}^{j}
    \right\|_2^2,
    \qquad
    j\in\{a,f,v\}.
\end{equation}

\subsubsection{Transformer-based Predictive Learner}

Each residual vector is first projected into a latent embedding space,
\begin{equation}
    e_k = W_e r_k+b_e+p_k,
\end{equation}
where $W_e$ and $b_e$ are learnable projection parameters and
$p_k$ is a learnable positional embedding associated with the
$k$th position in the history window.

For each Transformer layer, the embedded sequence is mapped to query,
key, and value representations,
\begin{equation}
    Q=EW_Q,
    K=EW_K,
    V=EW_V,
\end{equation}
where $E=[e_1,\ldots,e_H]$ denotes the embedded residual history.

The scaled dot-product self-attention operation is
\begin{equation}
    \operatorname{Attn}(Q,K,V)
    =
    \operatorname{softmax}
    \left(
        \frac{QK^\top}{\sqrt{d_k}}
    \right)V.
\end{equation}

Using $M$ attention heads, multi-head self-attention is given by
\begin{align}
    \mathrm{head}_m
    &=
    \operatorname{Attn}
    \left(
        Q_m,K_m,V_m
    \right),\\
    \operatorname{MHA}(E)
    &=
    \operatorname{Concat}
    \left(
        \mathrm{head}_1,\ldots,\mathrm{head}_M
    \right)W_O.
\end{align}

Each Transformer encoder layer combines the self-attention operation
with a position-wise feedforward network,
\begin{equation}
    \operatorname{FFN}(h)
    =
    W_2\,\sigma(W_1h+b_1)+b_2,
\end{equation}
together with residual connections and layer normalization. After
$L$ Transformer layers, the encoded history is
\begin{equation}
    Z
    =
    \operatorname{TransformerEncoder}_{\theta}
    (\mathcal{W}_t)
    =
    [z_1,\ldots,z_H].
\end{equation}

The representation corresponding to the most recent residual sample
is selected as the history representation,
$
    z_t = z_H.
$

The latent representation $z_t$ therefore summarizes the residual
history through self-attention over the complete $H$-step input
sequence. The residual states are partitioned into angle, frequency, and
voltage-related components. Separate prediction heads map $z_t$
directly to the future $P$-step residual trajectories for such components and are learned via a weighted loss function similar to the other learners. Finally, the learned residual correction is combined with the
operator-model trajectory as
$
    \hat{x}_t
    =
    x_t^{\mathrm{op}}+\hat r_t.
$

\subsection{Residual Learning Performance}

\begin{figure*}[!t]
    \centering
    \includegraphics[width=0.32\linewidth]{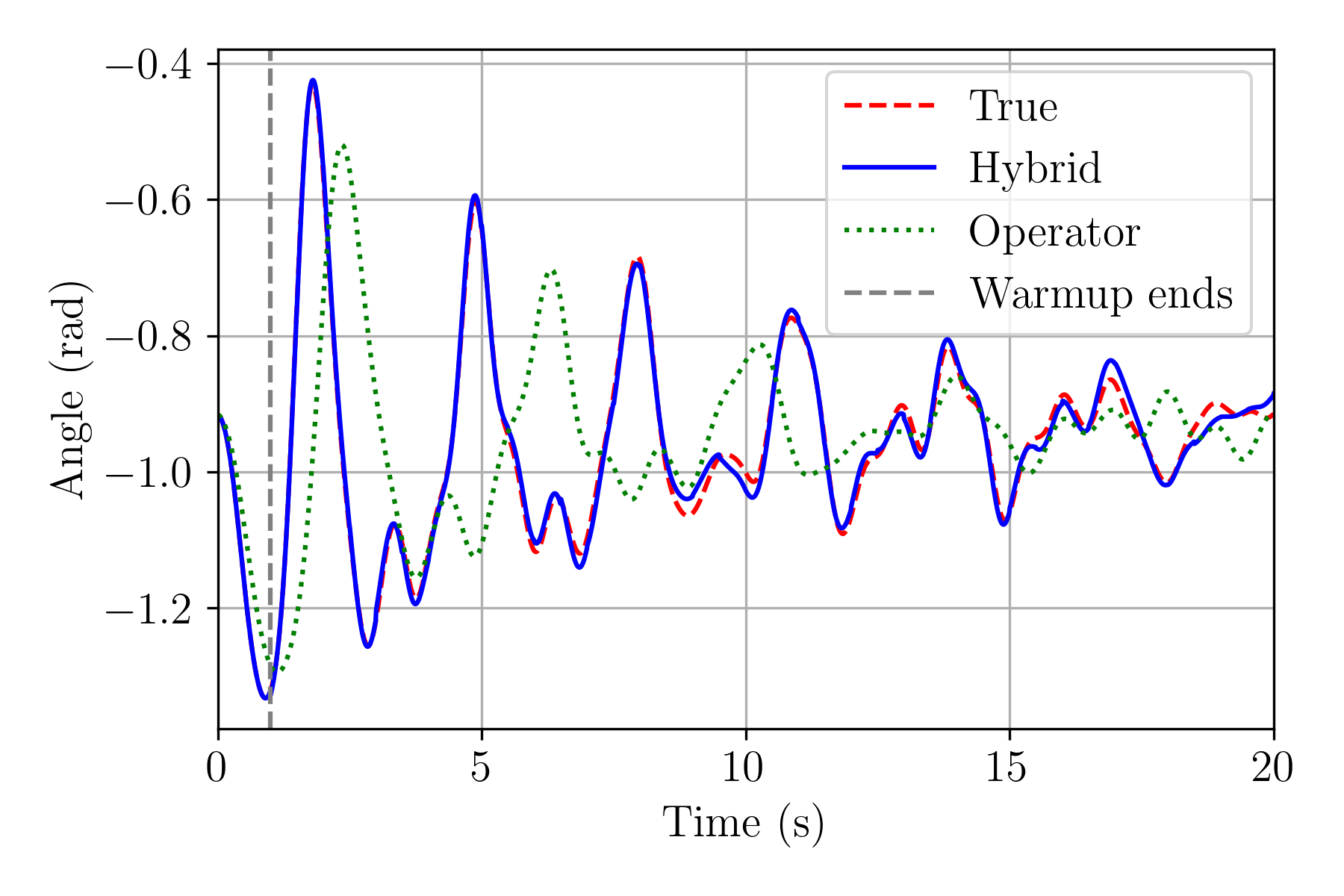}\hfill
    \includegraphics[width=0.32\linewidth]{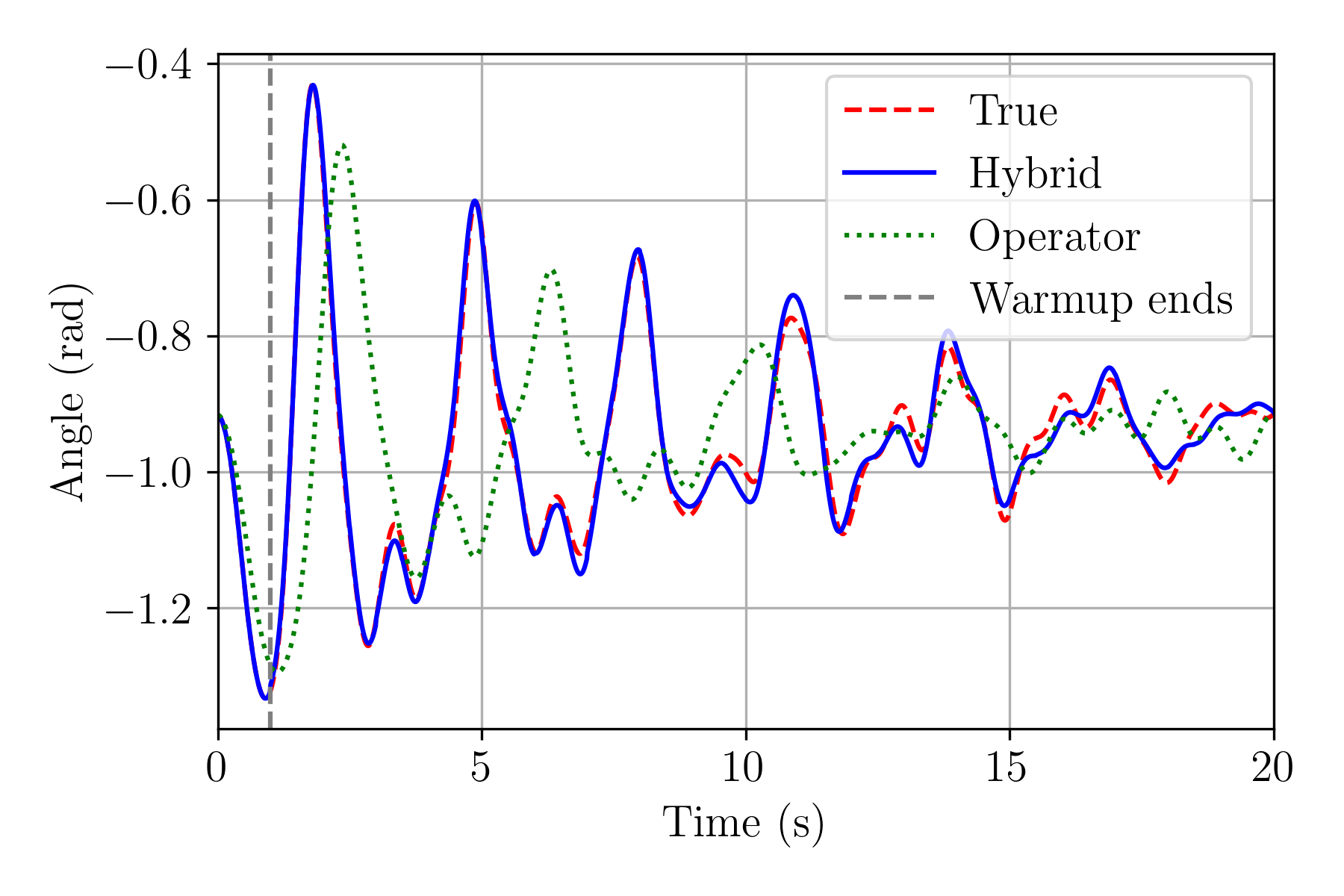}\hfill
    \includegraphics[width=0.32\linewidth]{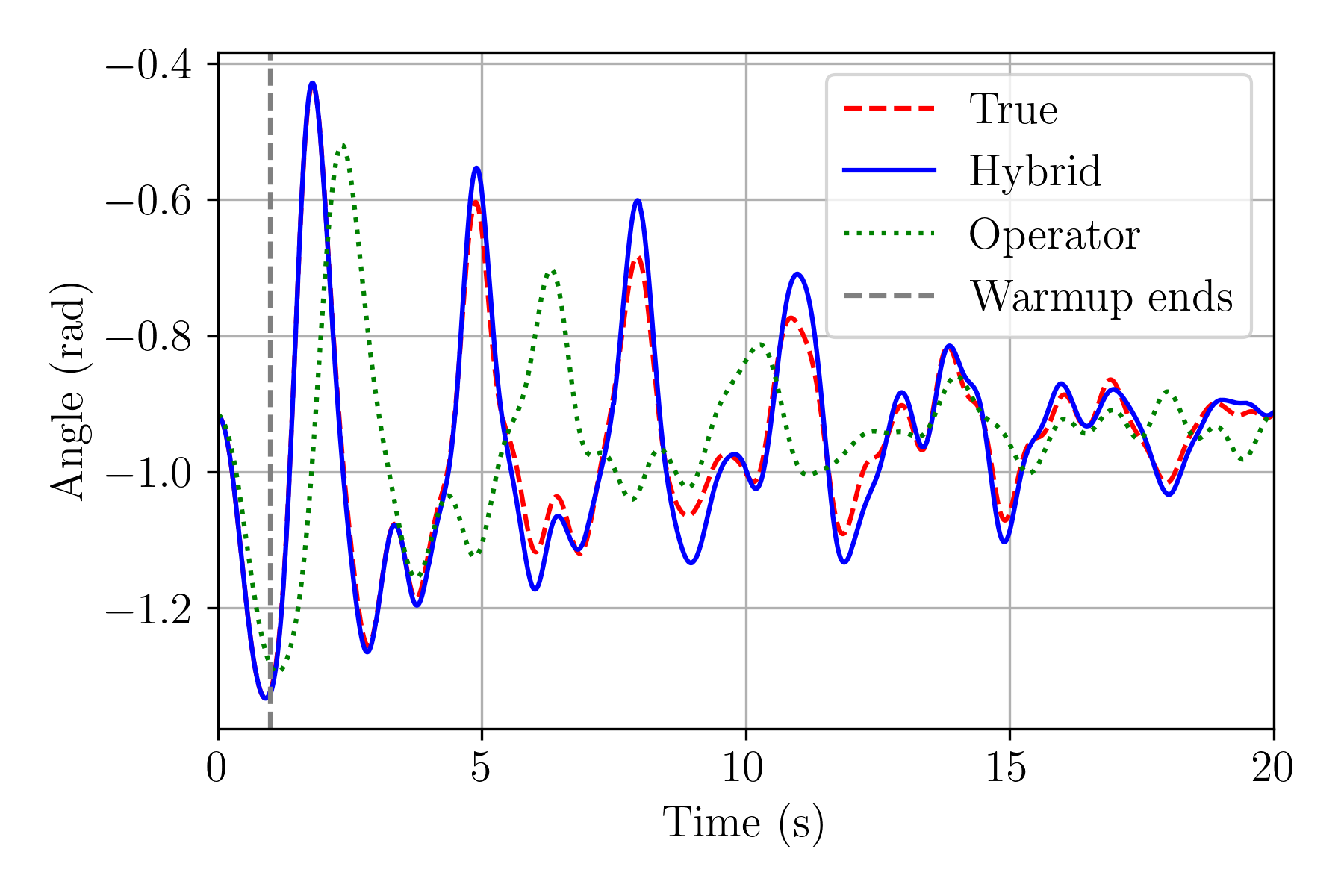}

    \includegraphics[width=0.32\linewidth]{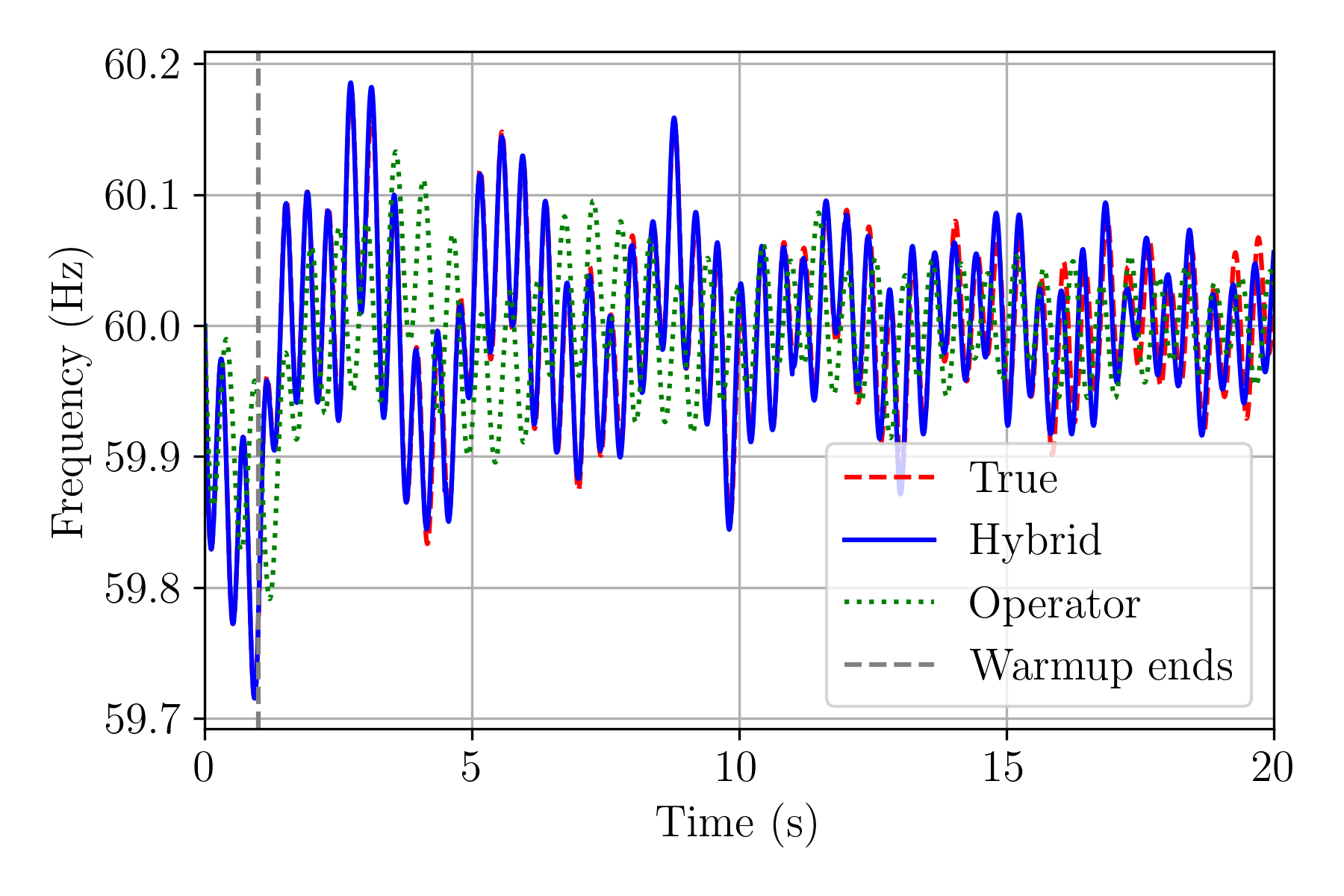}\hfill
    \includegraphics[width=0.32\linewidth]{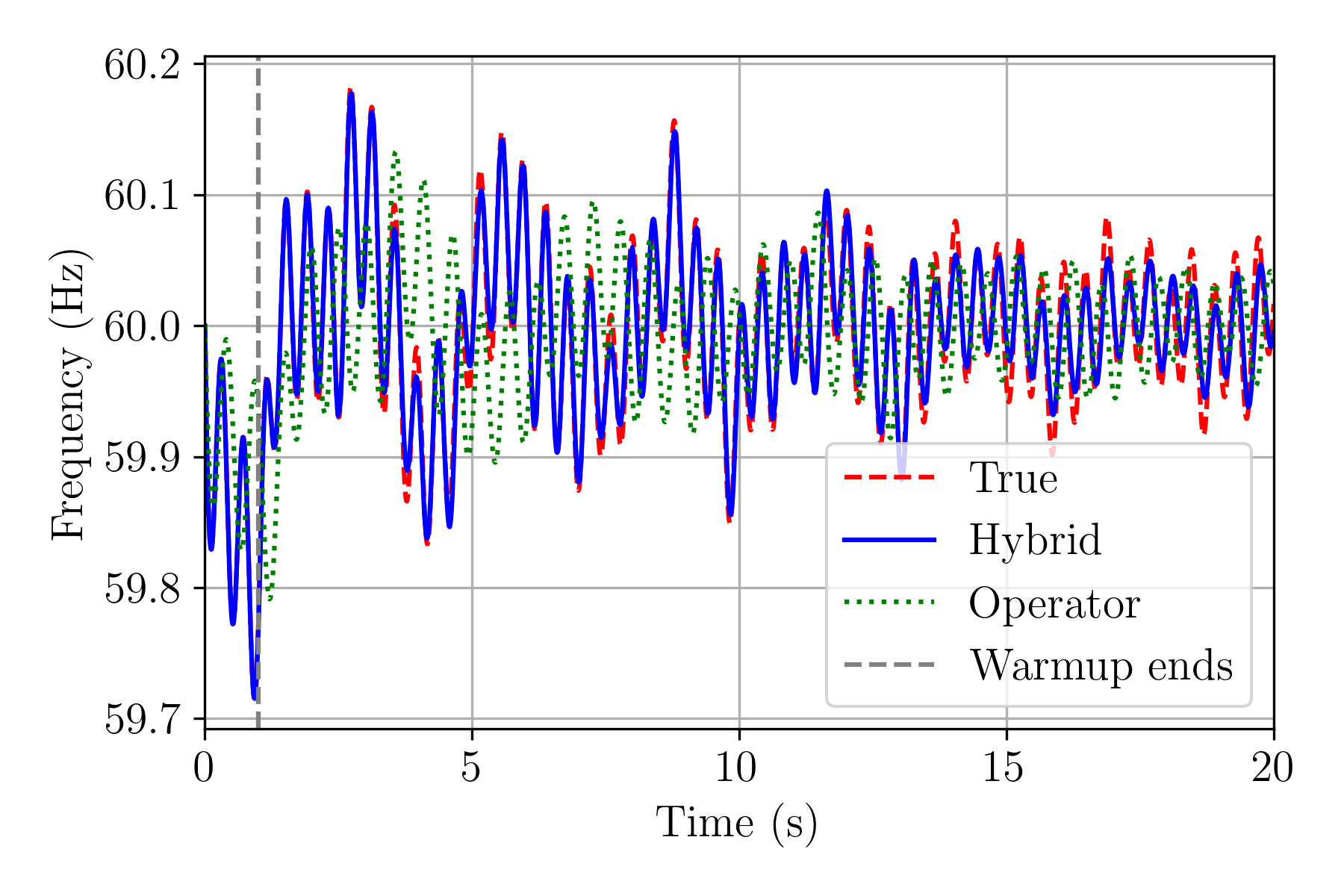}\hfill
    \includegraphics[width=0.32\linewidth]{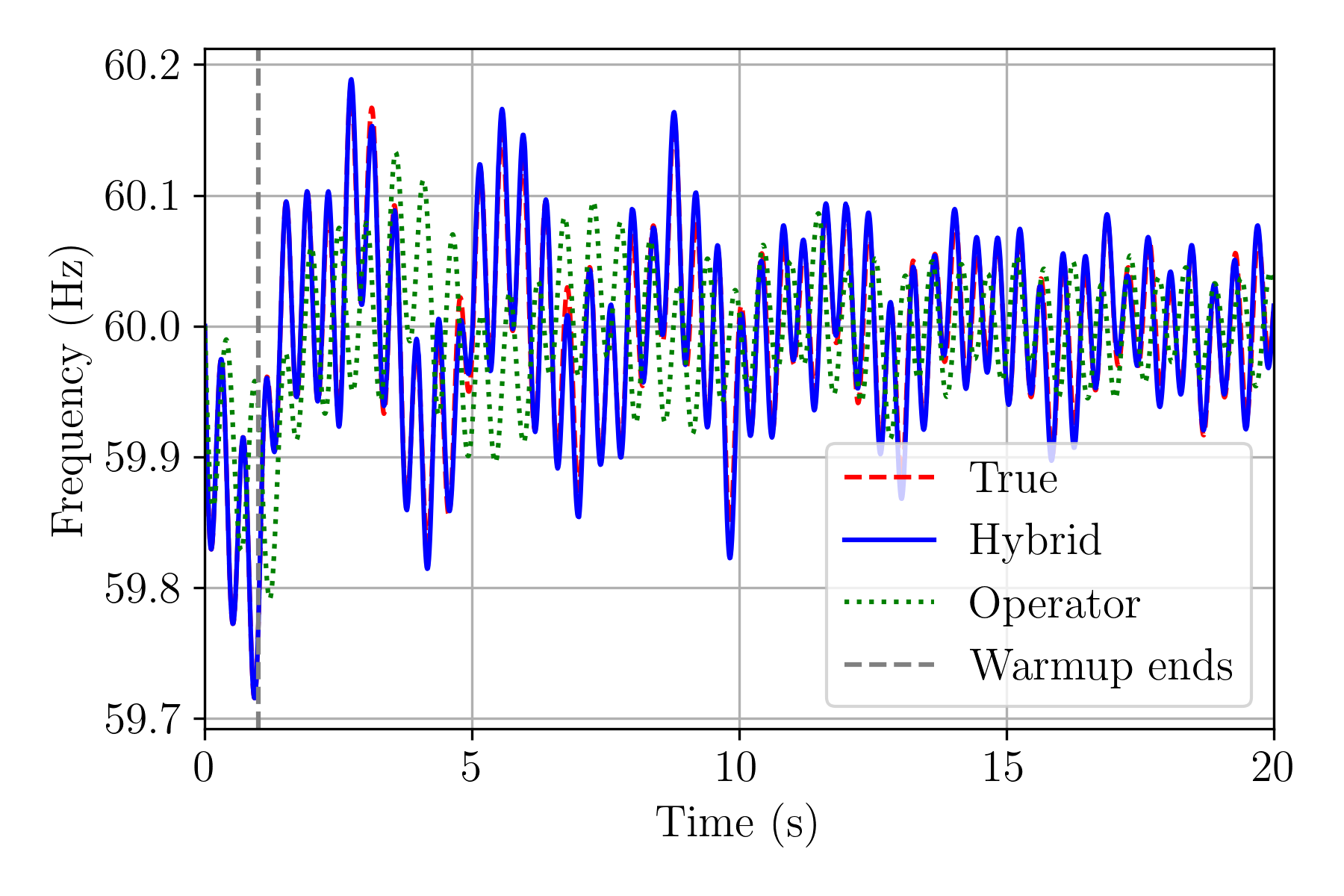}

    \includegraphics[width=0.32\linewidth]{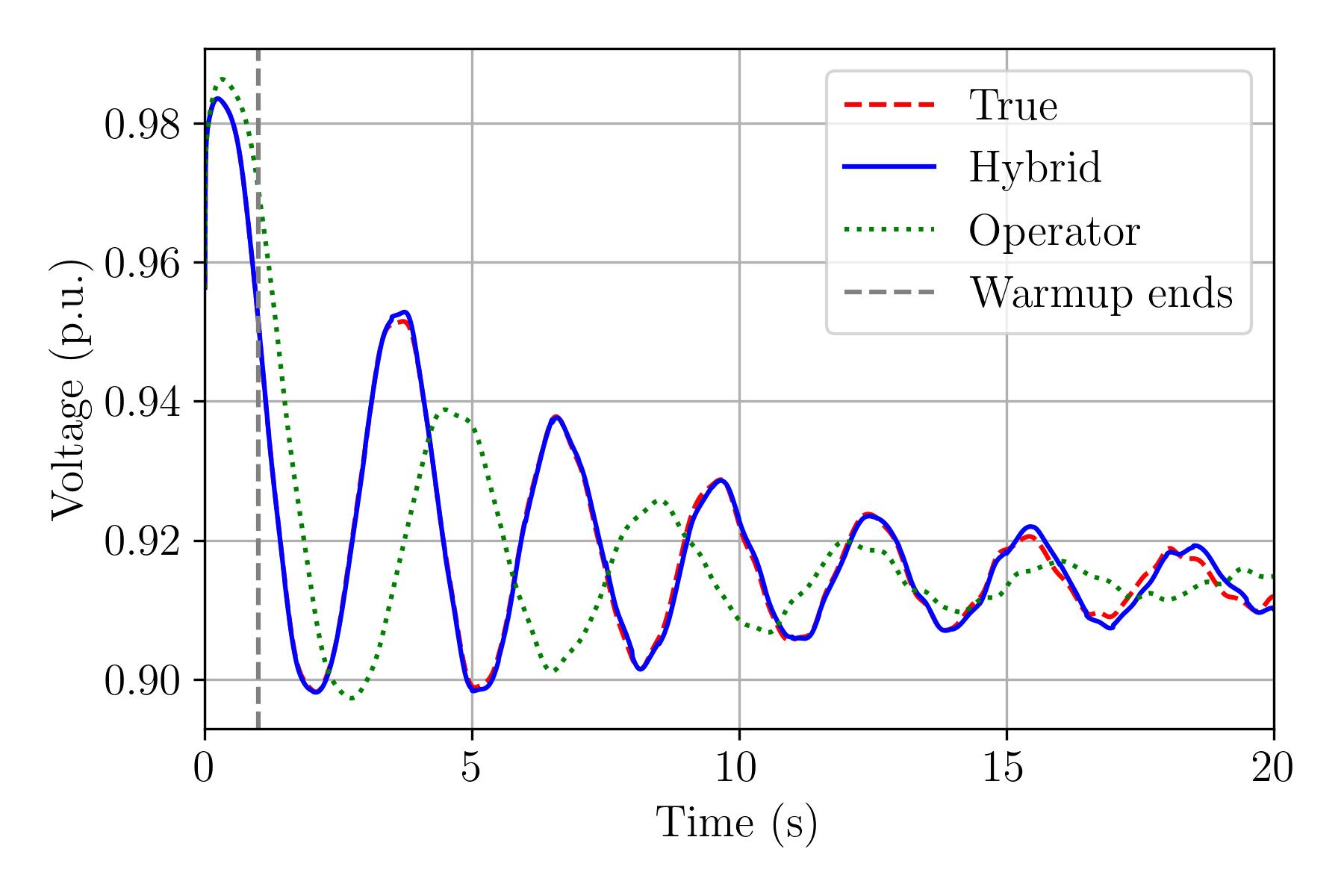}\hfill
    \includegraphics[width=0.32\linewidth]{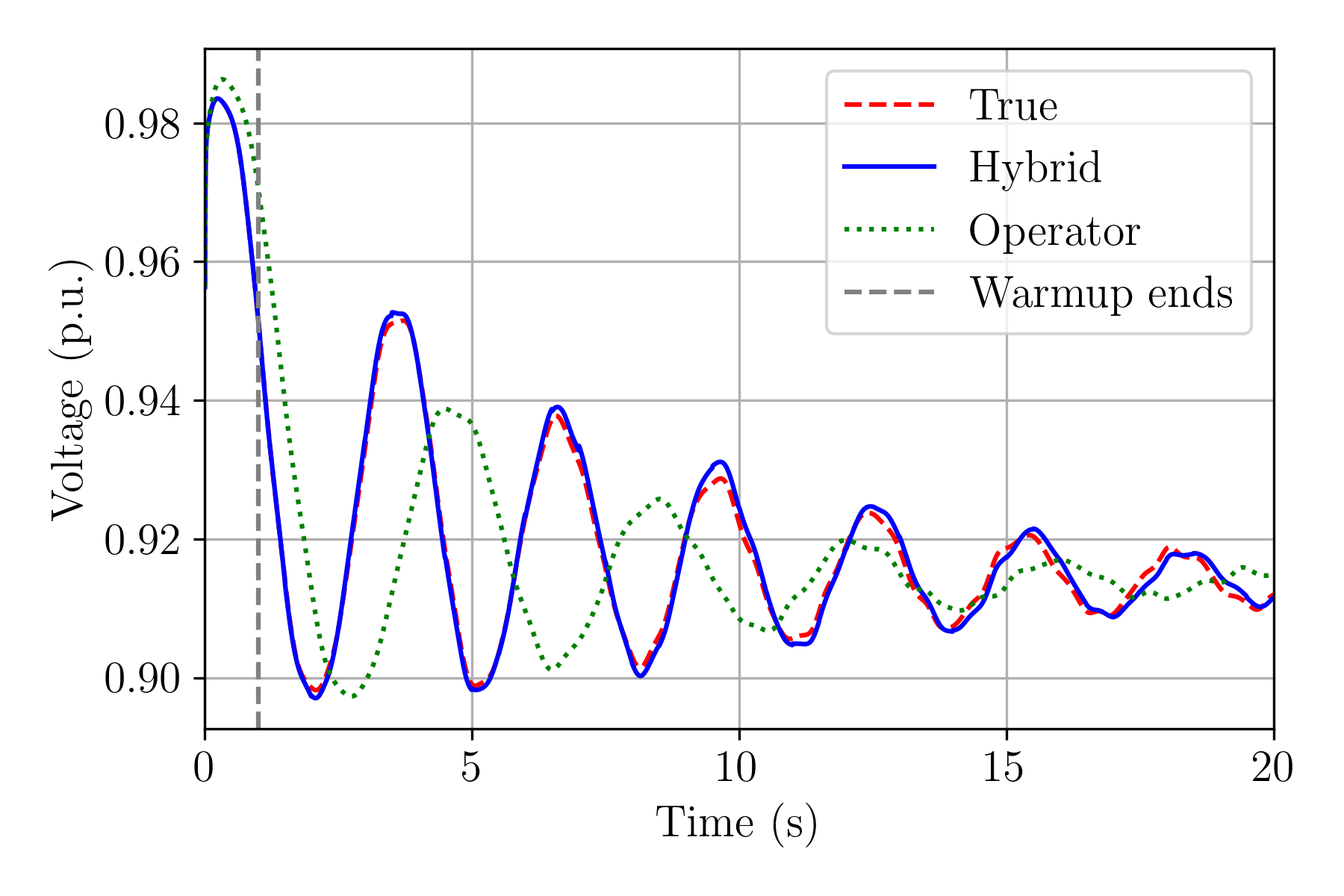}\hfill
    \includegraphics[width=0.32\linewidth]{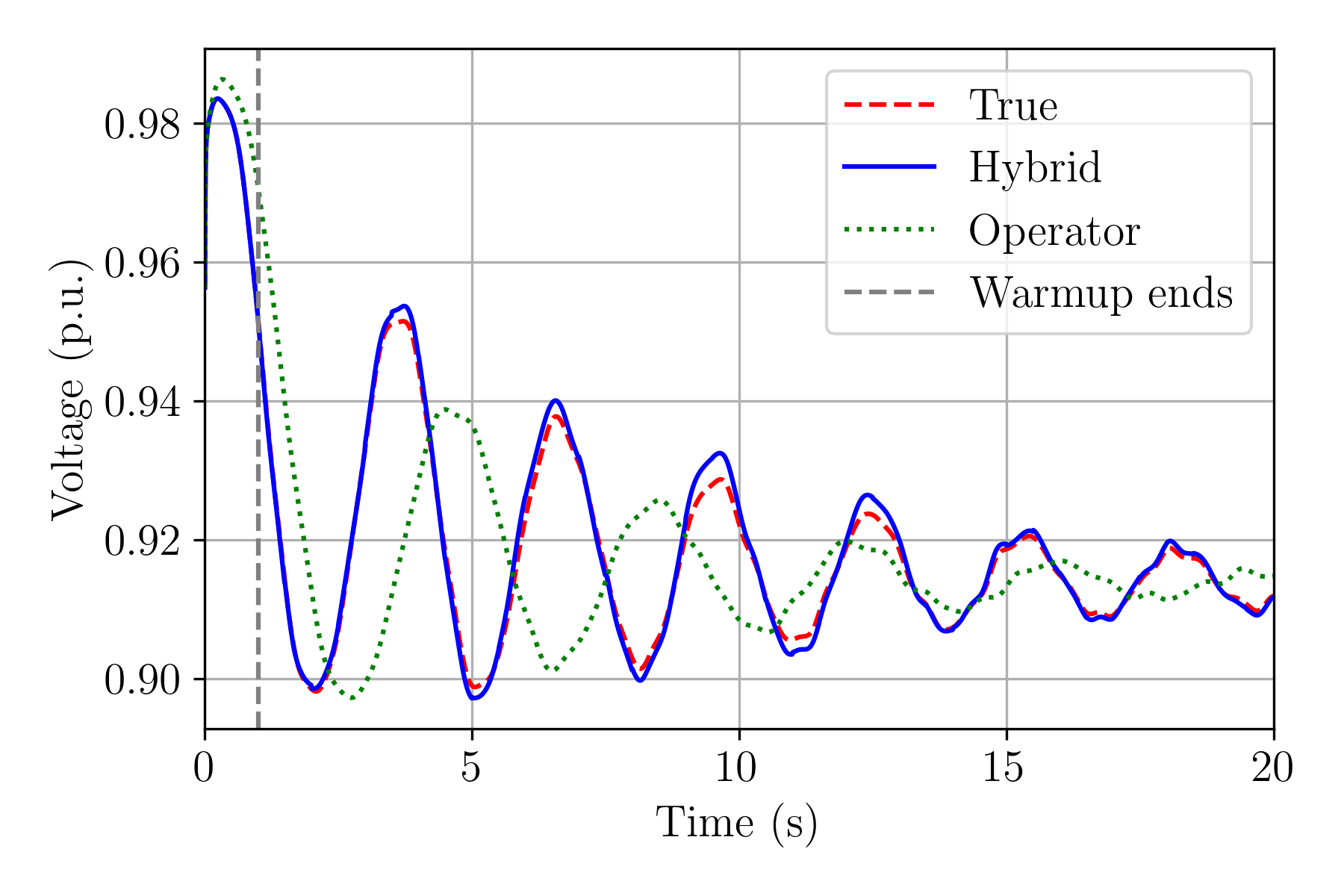}
    
    \caption{Comparison of predicted trajectories for grid dynamics. The rows correspond to angle (top), frequency (middle), and voltage (bottom) states. The columns represent the predictive models: LSTM (left), Neural ODE (center), and Transformer (right).}
    \label{fig:prediction_comparison}
\end{figure*}

\begin{table}[!t]
\centering
\caption{Hyperparameters and Network Configurations}
\label{tab:hyperparameters}
\begin{tabular}{ >{\raggedright\arraybackslash}p{1.5cm} >{\raggedright\arraybackslash}p{3.5cm} >{\centering\arraybackslash}p{2.3cm} }
\toprule
\textbf{Category} & \textbf{Parameter} & \textbf{Value} \\
\midrule
\multirow{8}{*}{\shortstack{\textbf{Common} \\ \textbf{Settings}}} 
 & History Window Length ($T_{hist}$) & 100 steps \\
 & Prediction Chunk Length ($T_{pred}$) & 50 steps \\
 & Hidden Size & 256 \\
 & Dropout Rate & 0.1 \\
 & Optimizer & Adam \\
 & Learning Rate & $3 \times 10^{-4}$ \\
 & Batch Size & 64 \\
 & Max Epochs & 900 \\
  & {Pre-training Trajectories} & {1000} \\
 & {Adaptation Trajectories} & {50} \\
\midrule
\multirow{2}{*}{\textbf{LSTM}} 
 & Architecture & Multi-layer LSTM \\
 & Number of Layers & 5 \\
\midrule
\multirow{3}{*}{\textbf{Neural ODE}}
 & Initial State Encoder & GRU (2 layers) \\
 & ODE Solver Method & Euler \\
 & ODE Function Network & 2-layer MLP (Tanh) \\
\midrule
\multirow{3}{*}{\textbf{Transformer}}
 & Number of Encoder Layers & 5 \\
 & Number of Attention Heads & 8 \\
 & Feedforward Dimension & 1024 \\
\bottomrule
\end{tabular}
\end{table}

\begin{table}[!t]
\centering
\caption{Prediction RMSE and Improvement on Test Set}
\label{tab:prediction_rmse}
\setlength{\tabcolsep}{4pt}
\resizebox{\columnwidth}{!}{%
\begin{tabular}{lccccccc}
\toprule
\multirow{2}{*}{\textbf{Method}} & \multirow{2}{*}{\textbf{Total}} & \multicolumn{2}{c}{\textbf{Angle (rad)}} & \multicolumn{2}{c}{\textbf{Freq. (Hz)}} & \multicolumn{2}{c}{\textbf{Volt. (p.u.)}} \\
\cmidrule(lr){3-4} \cmidrule(lr){5-6} \cmidrule(lr){7-8}
 & & \textbf{RMSE} & \textbf{Improv.} & \textbf{RMSE} & \textbf{Improv.} & \textbf{RMSE} & \textbf{Improv.} \\
\midrule
Operator & \color{gray} 0.07952 & \color{gray} 0.14656 & \color{gray} - & \color{gray} 0.06483 & \color{gray} - & \color{gray} 0.01365 & \color{gray} - \\
\midrule
LSTM & 0.01107 & 0.02043 & 86.06\% & 0.01317 & 79.69\% & 0.00147 & 89.21\% \\
Neural ODE & 0.01131 & 0.02087 & 85.76\% & 0.01659 & 74.41\% & 0.00154 & 88.70\% \\
Transformer & 0.00922 & 0.01701 & 88.39\% & 0.01231 & 81.01\% & 0.00135 & 90.08\% \\
\bottomrule
\end{tabular}%
}
\end{table}

To evaluate the effectiveness of the proposed residual learning framework, we compare the prediction performance of three state-of-the-art predictive learners: LSTM, Neural ODE, and Transformer models. 
The experiments are conducted on the inverter-integrated IEEE 68-bus system, where the operator model is subjected to parameter mismatches and unmodeled dynamics to simulate realistic discrepancies from the actual grid behavior.

The hyperparameter settings and network configurations for training these models are detailed in Table~\ref{tab:hyperparameters}. All predictive learners observe a historical trajectory window of $T_{hist} = 100$ steps to directly predict a future residual chunk of $T_{pred} = 50$ steps. The networks are trained using the Adam optimizer with an initial learning rate of $3 \times 10^{-4}$ for up to $900$ epochs.

Table~\ref{tab:prediction_rmse} presents a comparison of prediction accuracy between the uncorrected operator model and several residual-augmented models. Here, the reported RMSE for each state group represents the absolute error averaged over all corresponding states and across the entire prediction horizon, rather than a relative percentage error. The Total RMSE is a dimensionless quantity averaged across all normalized channels. The operator model yields a total RMSE of $0.07952$, highlighting the gap between its baseline dynamics and the actual system behavior. By integrating learned residual corrections, this error is significantly reduced across all three predictive learners: the Neural ODE, LSTM, and Transformer models achieve reductions of approximately $85.8\%$, $86.1\%$, and $88.4\%$, respectively. Among these, the Transformer-based residual learner delivers the lowest overall RMSE at $0.00922$, consistently outperforming the others across angle, frequency, and voltage state groups. Figure~\ref{fig:prediction_comparison} visually reinforces these improvements; while the uncorrected operator trajectories show discrepancies from the true dynamics, especially in oscillatory angle and voltage responses, the hybrid trajectories, enhanced by residual corrections, closely follow the actual system throughout the prediction horizon. Taken together, both quantitative and trajectory-level evidence demonstrate that learning and correcting the discrepancy, rather than replacing the operator model, effectively compensates for baseline inaccuracies in grid dynamics.

\subsection{Adaptation Performance}

\begin{figure*}[!t]
    \centering
    \includegraphics[width=0.32\linewidth]{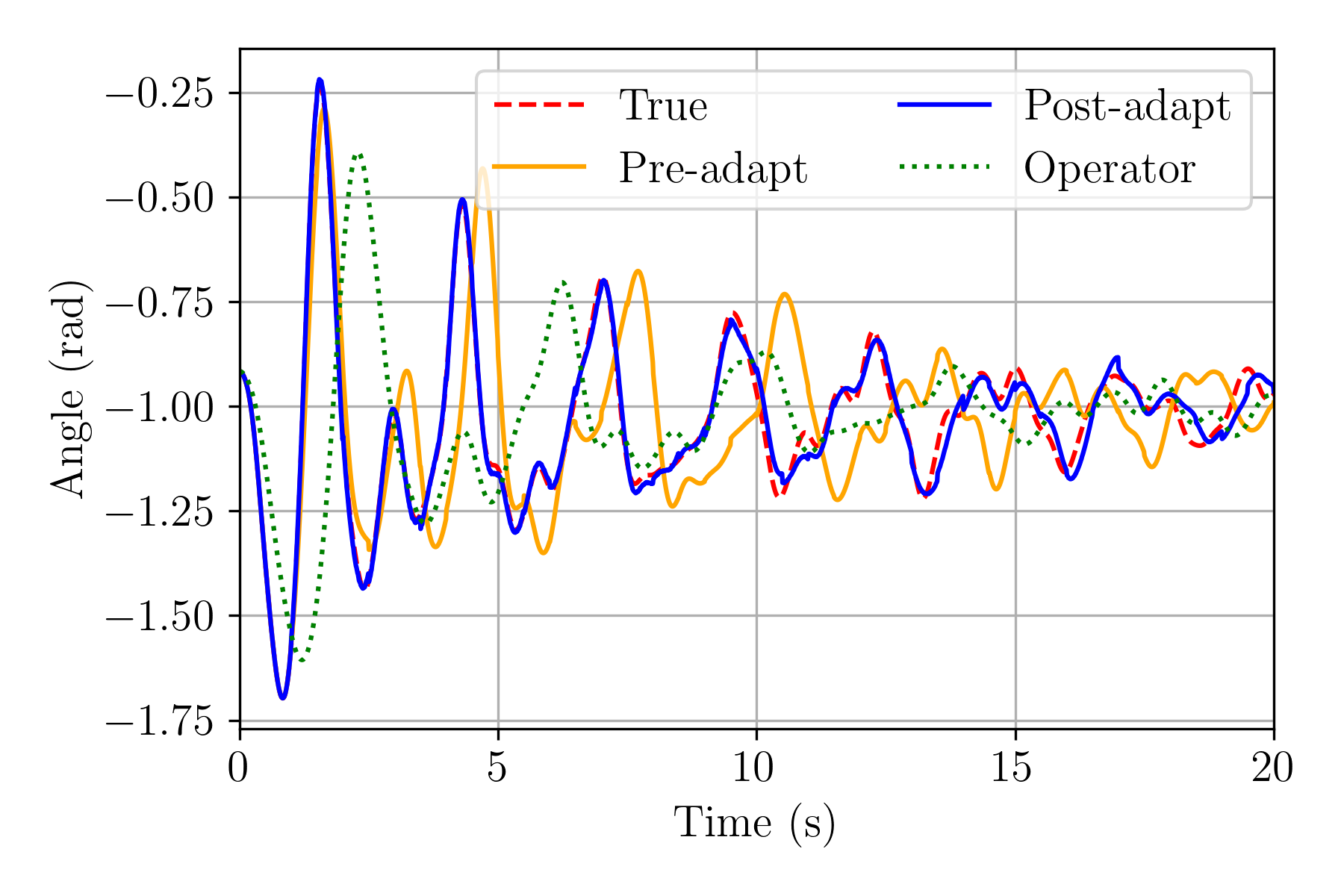}\hfill
    \includegraphics[width=0.32\linewidth]{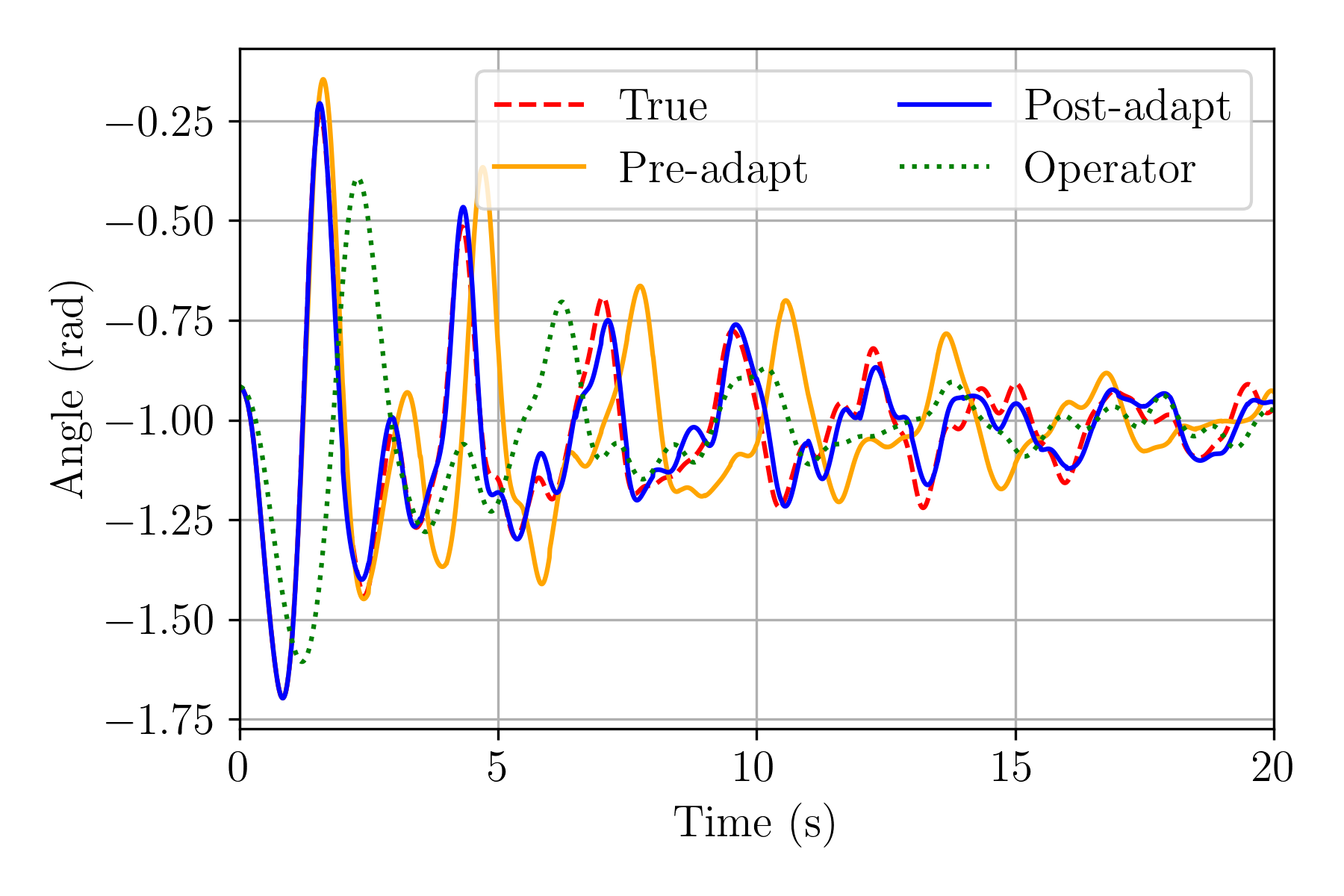}\hfill
    \includegraphics[width=0.32\linewidth]{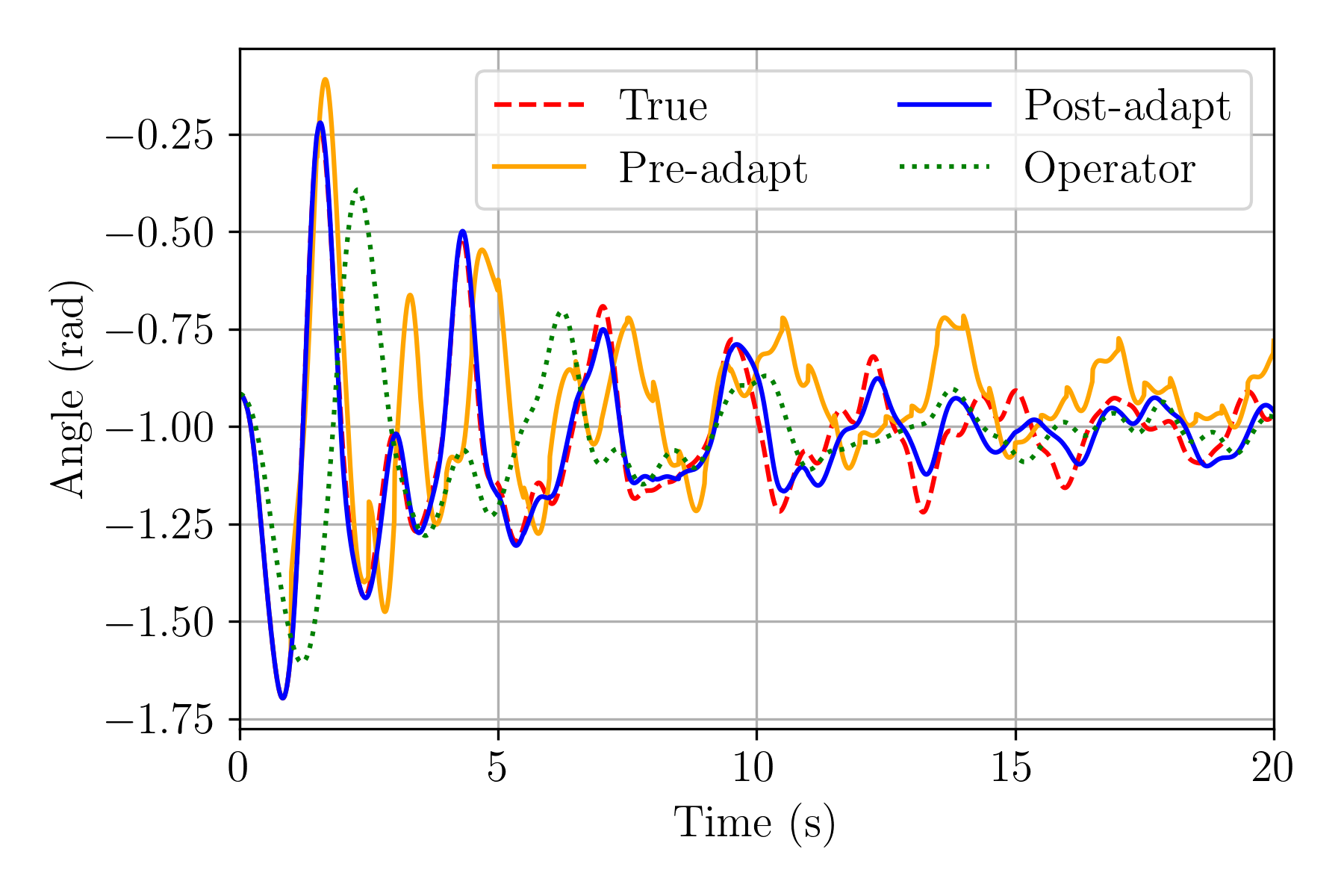}

    \includegraphics[width=0.32\linewidth]{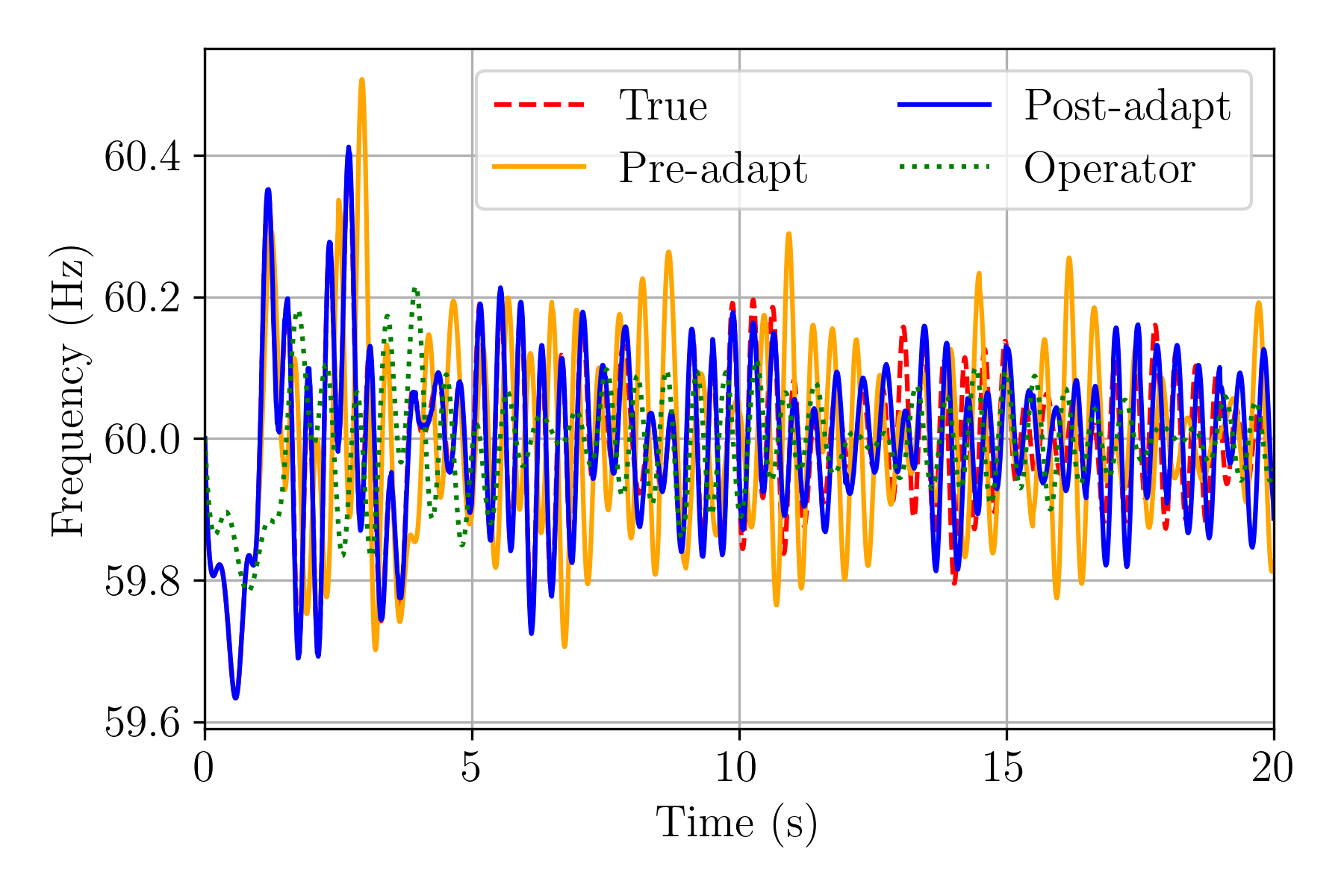}\hfill
    \includegraphics[width=0.32\linewidth]{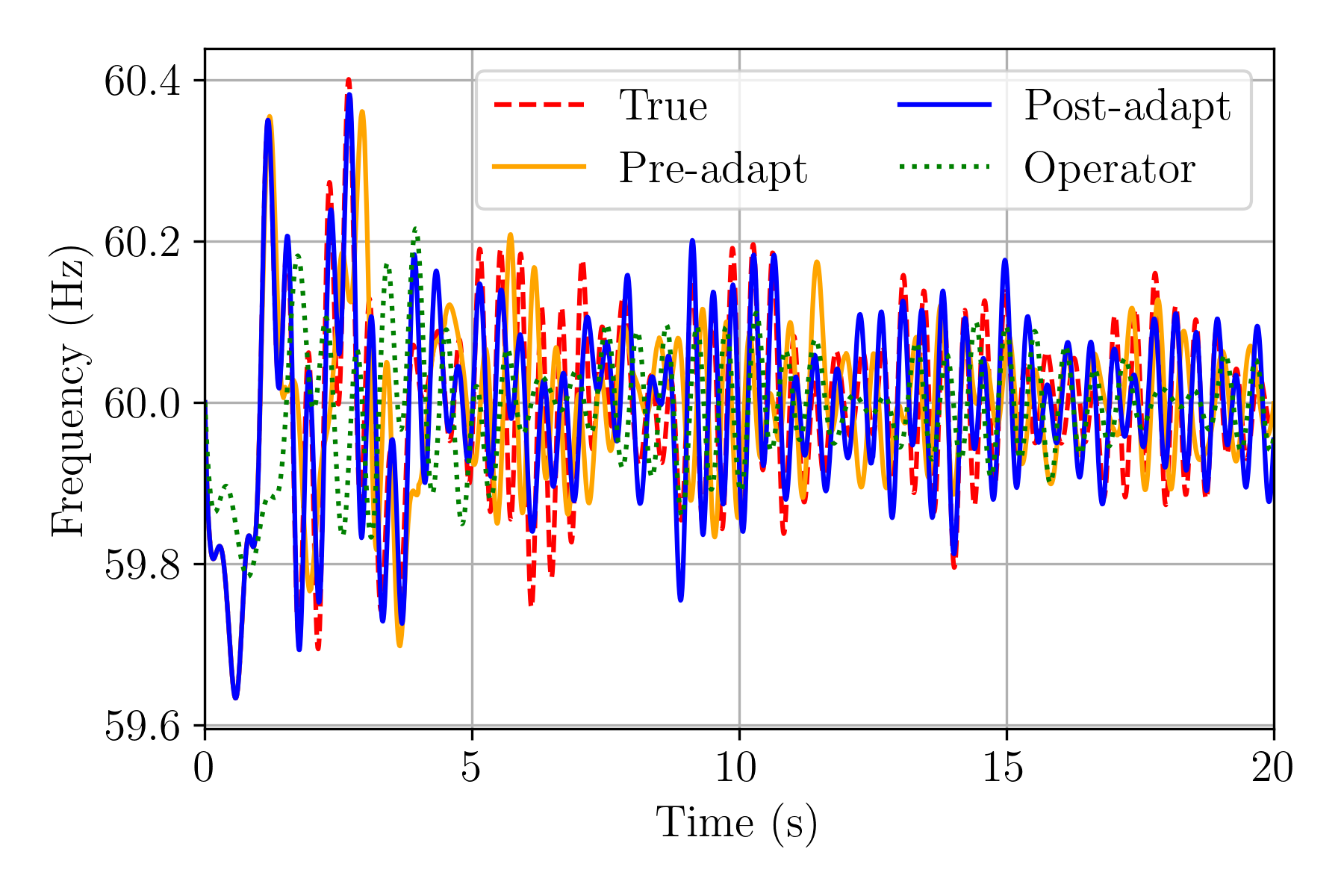}\hfill
    \includegraphics[width=0.32\linewidth]{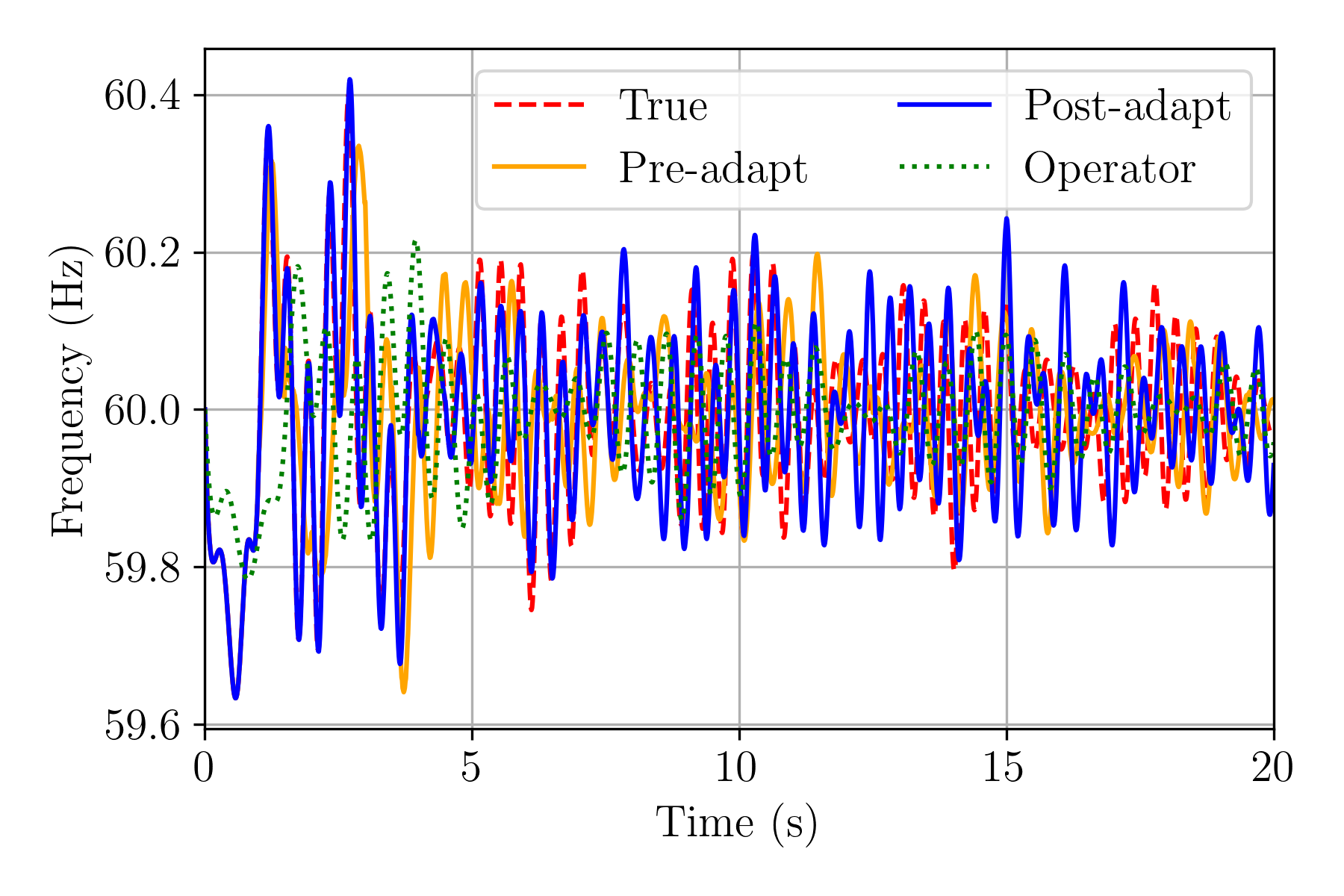}

    \includegraphics[width=0.32\linewidth]{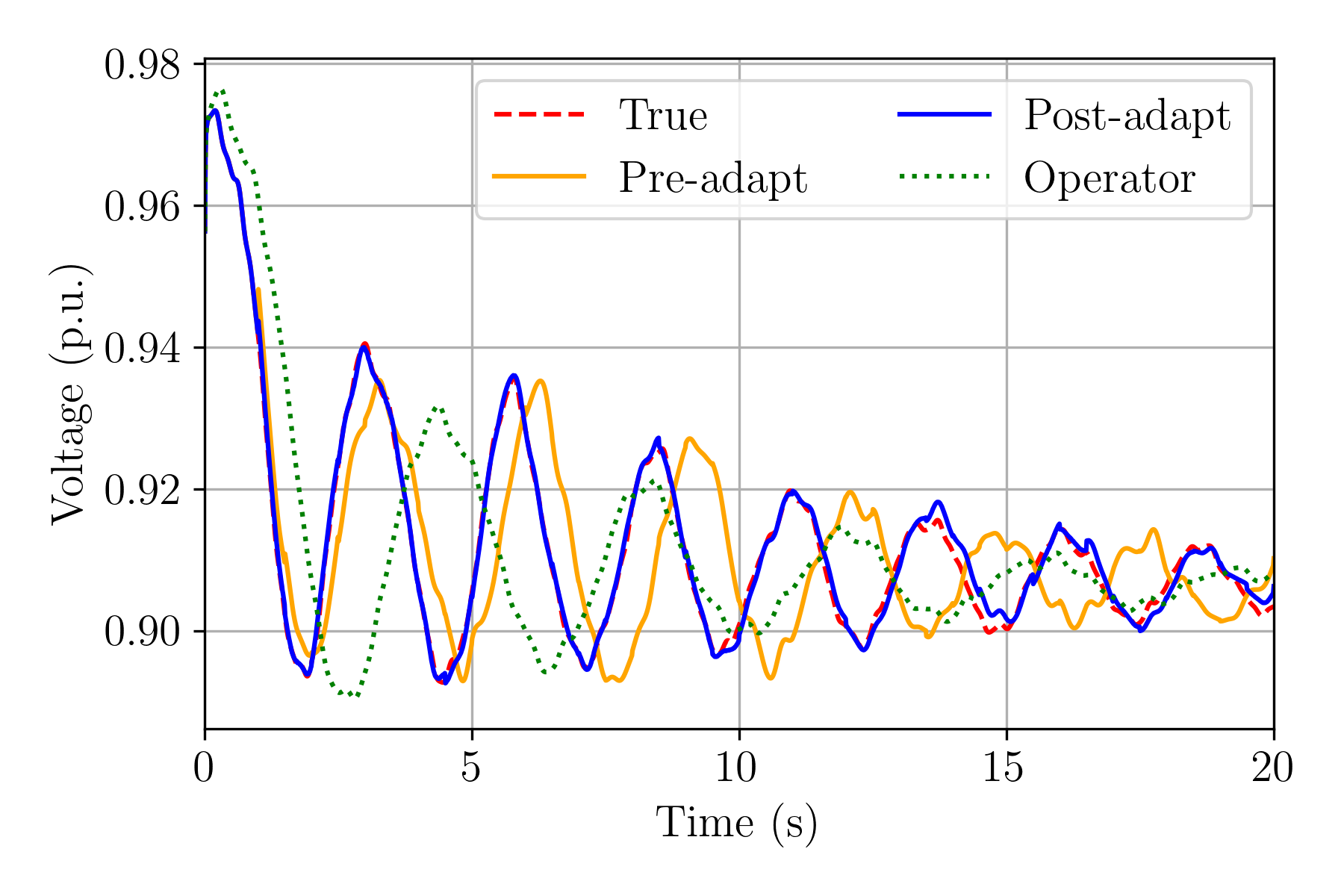}\hfill
    \includegraphics[width=0.32\linewidth]{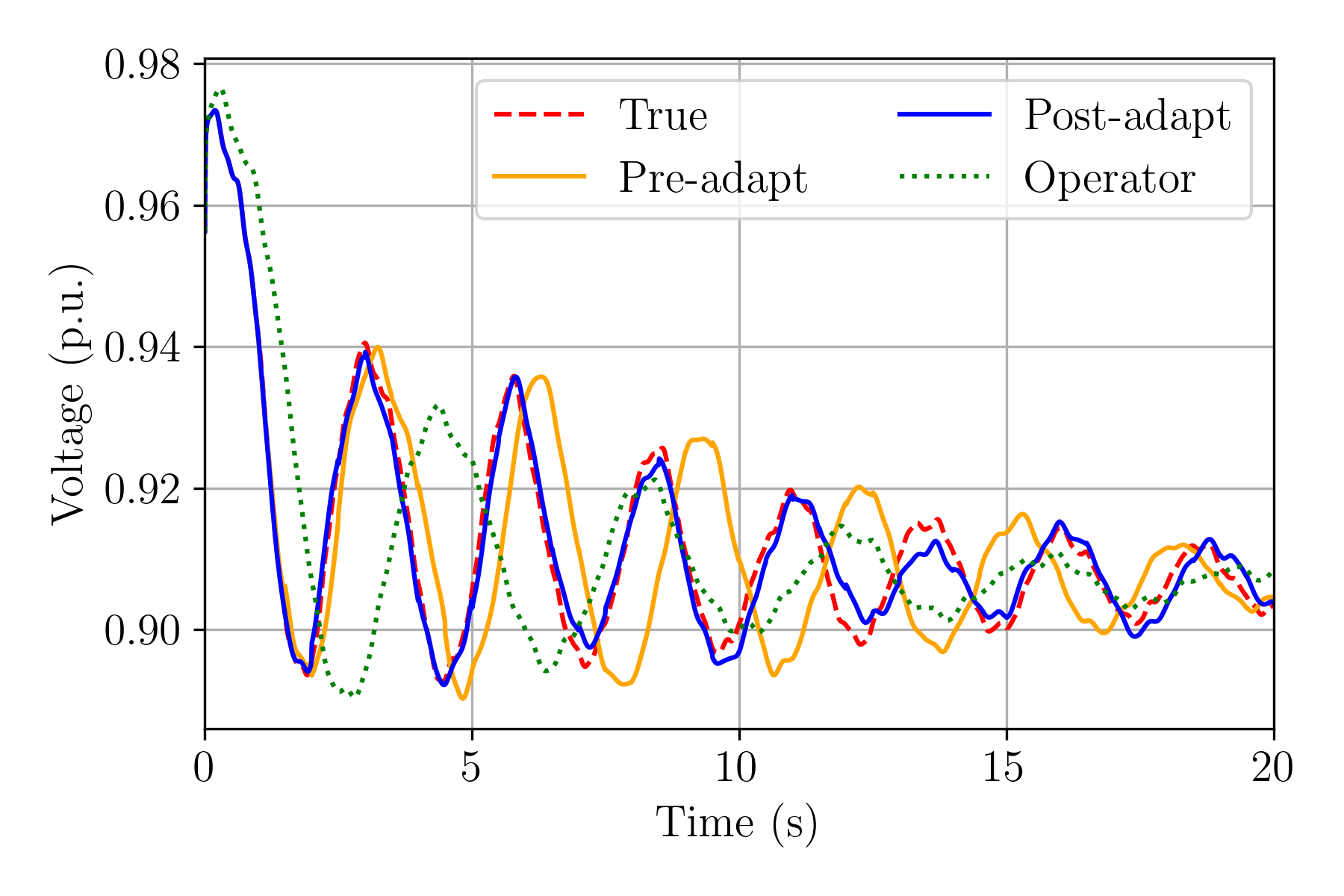}\hfill
    \includegraphics[width=0.32\linewidth]{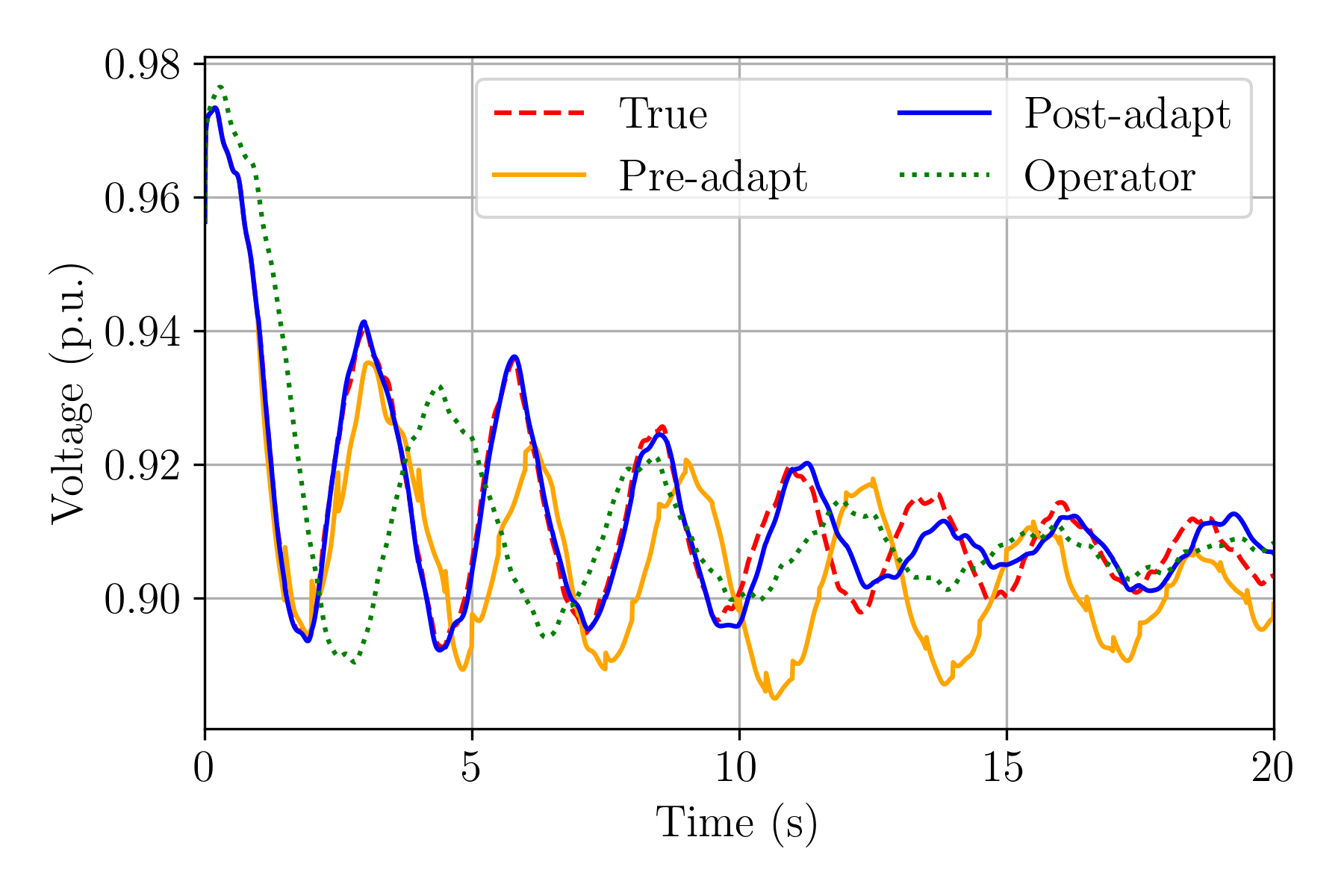}
    
    \caption{Comparison of predicted trajectories for grid dynamics after few-shot adaptation on the new grid system. The rows correspond to angle (top), frequency (middle), and voltage (bottom) states. The columns represent the predictive models: LSTM (left), Neural ODE (center), and Transformer (right).}
    \label{fig:adaptation_comparison}
\end{figure*}
\begin{figure*}[!t]
    \centering
    \includegraphics[width=0.32\linewidth]{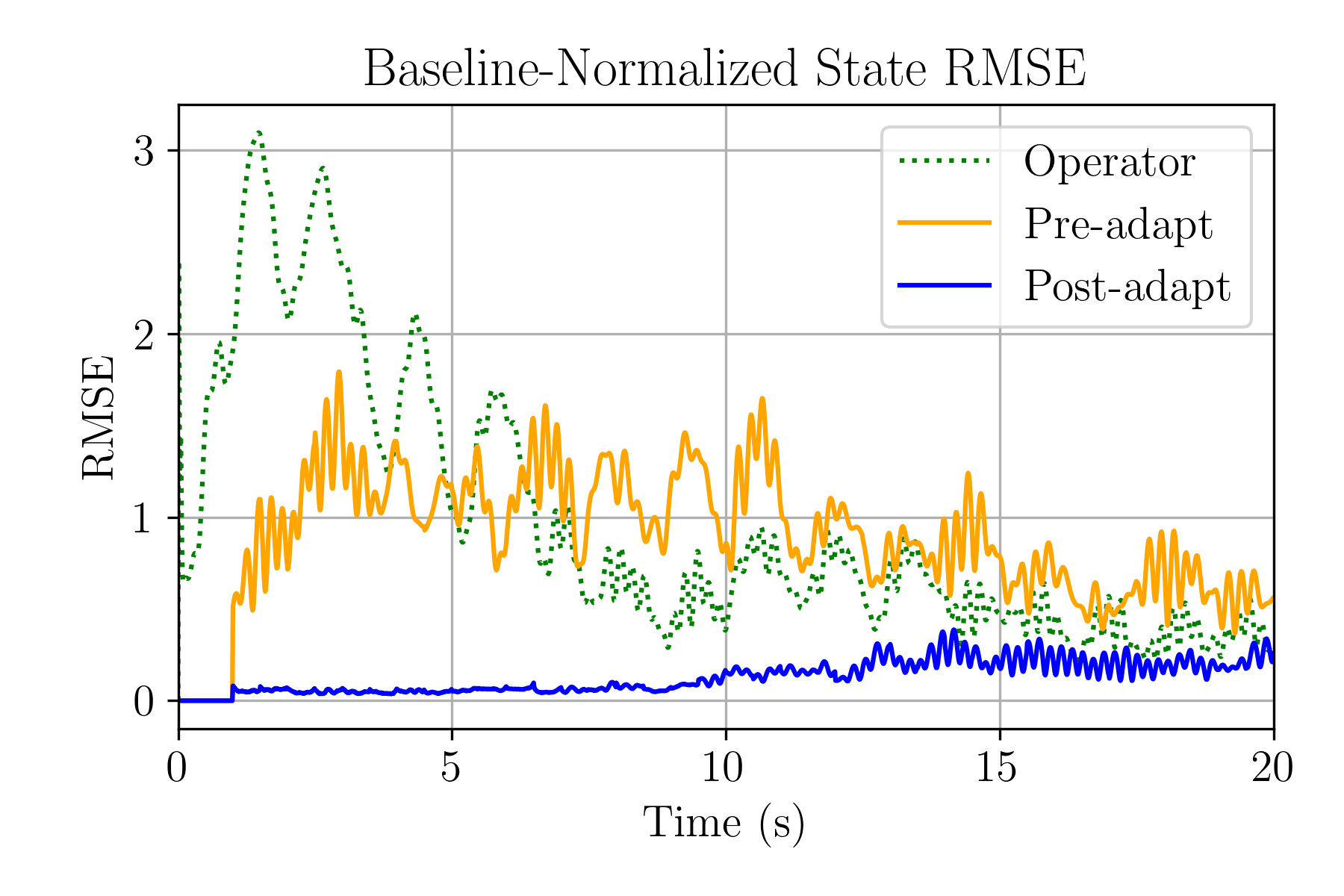}\hfill
    \includegraphics[width=0.32\linewidth]{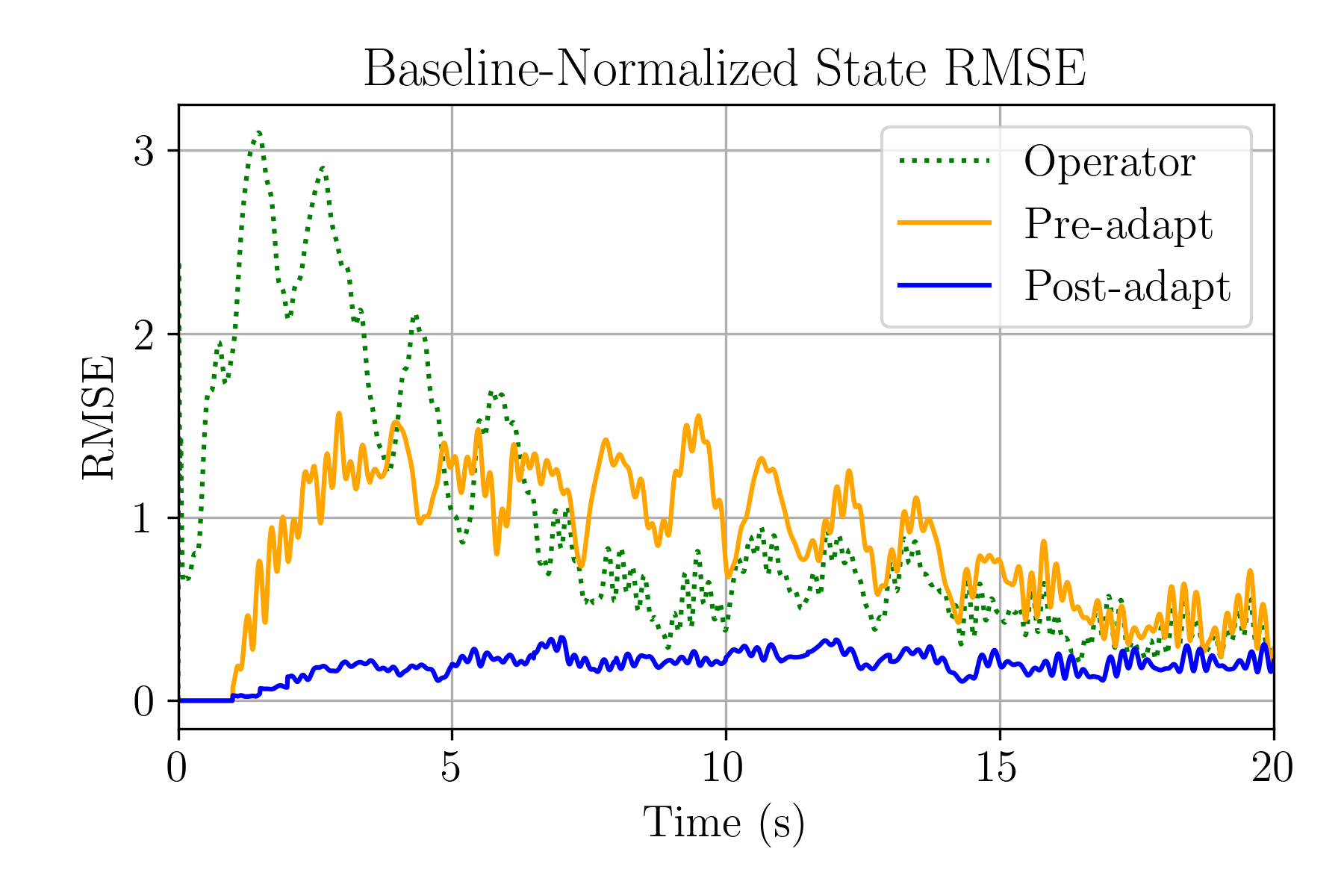}\hfill
    \includegraphics[width=0.32\linewidth]{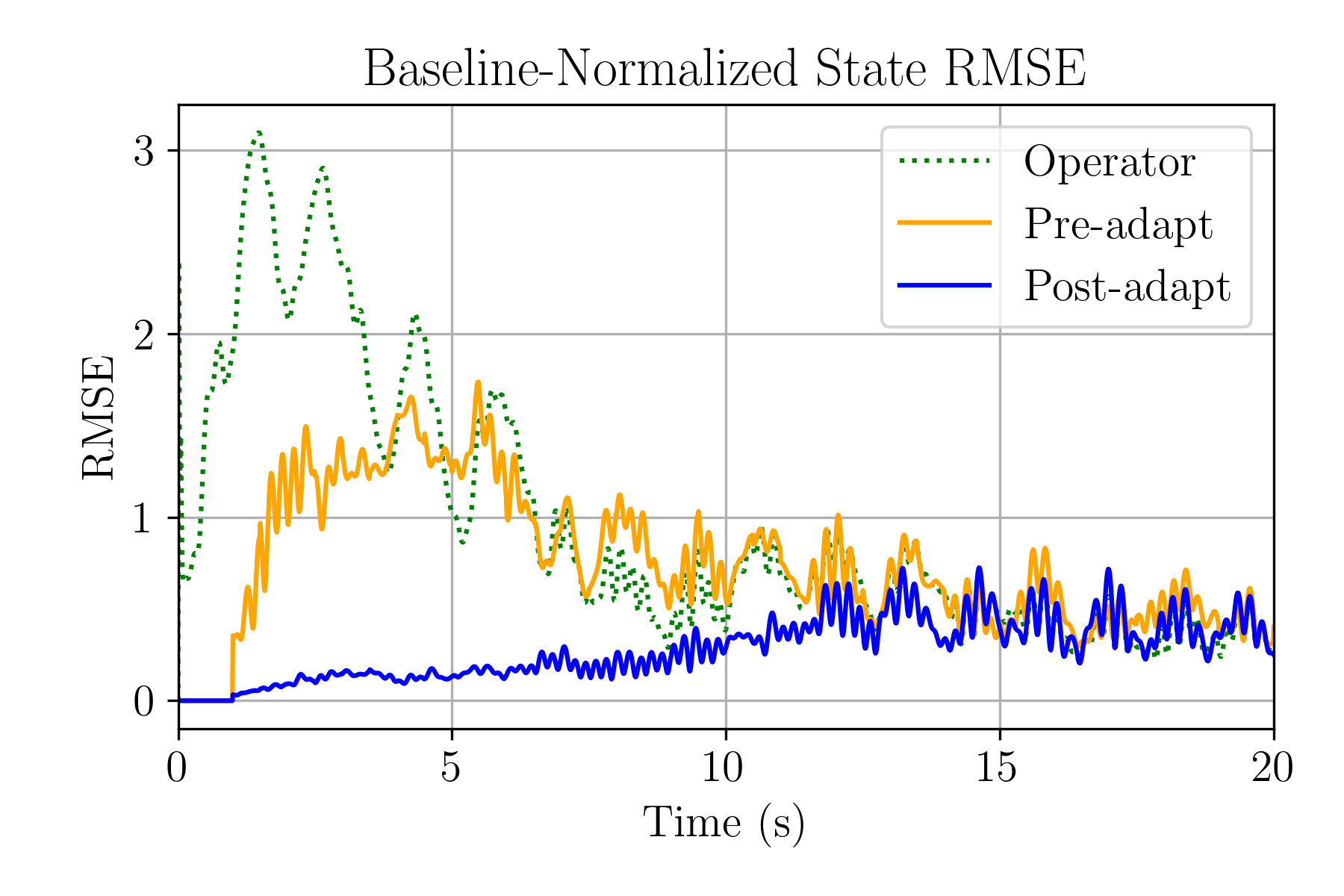}
    \caption{Comparison of baseline-normalized aggregated state RMSE over time on the adaptation dataset. The squared errors of angle, frequency, and voltage are weighted by the inverse of their respective operator MSEs to prevent magnitude domination. The plots demonstrate the prediction error of the baseline Operator and the predictive models before (Pre-adapt) and after (Post-adapt) few-shot training: LSTM (left), Neural ODE (center), and Transformer (right).}
    \label{fig:adaptation_state_rmse}
\end{figure*}

To validate the adaptability of the trained models to operational changes, we evaluate the proposed few-shot adaptation strategy on a modified grid configuration (System~3). In this scenario, the grid undergoes structural or parameter shifts, and only a limited amount of new trajectory data is available. To prevent catastrophic forgetting while rapidly adapting to the new dynamics, we employ an architecture-specific partial freezing strategy combined with a replay buffer of the original data. Rather than fine-tuning the entire network, we freeze the lower layers of the encoders to preserve low-level temporal features and train only the upper representation layers and prediction heads. Specifically, for the LSTM, we freeze the first recurrent layer and update the second layer alongside the output heads. Similarly, for the Neural ODE, the first GRU layer is frozen while the second GRU layer and the ODE function are updated. For the Transformer, which is particularly prone to overfitting on scarce data, initial adaptation attempts revealed a critical difficulty: standard partial fine-tuning failed to capture the altered oscillation frequencies of the new grid topology, as the model remained constrained by its pre-trained temporal references. To overcome this bottleneck, we employ a targeted parameter-efficient adapter strategy. We freeze the entire pre-trained encoder (layers 0--4) and the input projections to perfectly preserve the learned foundational physics. To enable adaptation to the new grid's shifted frequency and dynamics, we add a trainable 2-layer MLP bottleneck adapter directly after the encoder, and critically, we unfreeze the positional embeddings to allow the time-scale reference to recalibrate. The linear heads are also fine-tuned. This ensures that the model adapts flexibly without catastrophic forgetting.

\begin{table}[!t]
\centering
\caption{Prediction RMSE and Improvement on the Adaptation Dataset}
\label{tab:adaptation_rmse_comparison}
\setlength{\tabcolsep}{4pt}
\resizebox{\columnwidth}{!}{%
\begin{tabular}{lccccccc}
\toprule
\multirow{2}{*}{\textbf{Method}} & \multirow{2}{*}{\textbf{Total}} & \multicolumn{2}{c}{\textbf{Angle (rad)}} & \multicolumn{2}{c}{\textbf{Freq. (Hz)}} & \multicolumn{2}{c}{\textbf{Volt. (p.u.)}} \\
\cmidrule(lr){3-4} \cmidrule(lr){5-6} \cmidrule(lr){7-8}
 & & \textbf{RMSE} & \textbf{Improv.} & \textbf{RMSE} & \textbf{Improv.} & \textbf{RMSE} & \textbf{Improv.} \\
\midrule
\color{gray} Operator & \color{gray} 0.09610 & \color{gray} 0.17691 & \color{gray} - & \color{gray} 0.10111 & \color{gray} - & \color{gray} 0.01424 & \color{gray} - \\
\midrule
LSTM & 0.01482 & 0.02735 & 84.54\% & 0.03999 & 60.45\% & 0.00141 & 90.10\% \\
Neural ODE & 0.01609 & 0.02964 & 83.25\% & 0.02731 & 72.99\% & 0.00220 & 84.55\% \\
{Transformer} & {0.02255} & {0.04154} & {76.51\%} & {0.04620} & {54.43\%} & {0.00304} & {78.65\%} \\
\bottomrule
\end{tabular}%
}
\end{table}

The qualitative results of this few-shot adaptation are shown in Fig.~\ref{fig:adaptation_comparison}. The operator model's predictions significantly diverge from the actual dynamics of the new system, yielding substantial errors across all states. Following the adaptation phase, both the LSTM and Neural ODE models successfully recalibrate to the new system dynamics, maintaining tight tracking of the actual angle, frequency, and voltage trajectories. The Transformer model, utilizing the
proposed adapter strategy and unfrozen positional embeddings, also successfully captures the shifted dynamics, tracking the actual trajectories closely and demonstrating strong few-shot adaptability without overfitting.

To prevent the numerically larger variables from dominating the overall error metric, we calculate a baseline-normalized Root Mean Square Error (RMSE). Specifically, the squared errors of the angle, frequency, and voltage states are independently normalized using the inverse of their respective uncorrected operator MSE as weights. This ensures that the overall RMSE fairly represents the relative improvement across all physical domains.
To illuminate the adaptation process, Fig.~\ref{fig:adaptation_state_rmse} illustrates the temporal evolution of this baseline-normalized RMSE for each architecture. By comparing the error trajectories of the initially pre-trained model (Pre-adapt) with the fine-tuned model (Post-adapt) and the baseline Operator, the efficacy of the few-shot learning phase becomes highly apparent. Across all architectures, the Post-adapt curves consistently remain below both the Operator and Pre-adapt curves throughout the 20-second window, demonstrating that the adaptation successfully corrects the performance degradation caused by the system shift. Notably, while the LSTM and Neural ODE rapidly stabilize at a low error magnitude, the Transformer curve (Post-adapt) now also achieves a comparably low and stable error profile. This completely overcomes the high error variance seen in its pre-trained counterpart, visually corroborating the effectiveness of the targeted adapter-based tuning in adjusting to the new domain with restricted samples.
Table~\ref{tab:adaptation_rmse_comparison} provides a quantitative breakdown of the adaptation performance. The uncorrected operator model exhibits a high total RMSE of 0.09610. Among the adapted models, the LSTM achieves the best overall performance with a total RMSE of 0.01482, effectively reducing the angle error by 84.54\% and the voltage error by 90.10\%. The Neural ODE also demonstrates strong few-shot adaptability, yielding a total RMSE of 0.01609 and the highest frequency improvement (72.99\%). The Transformer model, utilizing the adapter strategy, records a competitive total RMSE of 0.02255, with substantial improvement margins across all states (e.g., 76.51\% for angle and 78.65\% for voltage).

\section{Concluding Remarks}

This work demonstrates that supplementing traditional operator simulation models with AI-based residuals offers an effective solution to the growing discrepancies between modeled and actual grid dynamics in modern power systems. By leveraging sensing data, our hybrid framework maintains nominal physics-based information while dynamically adapting to evolving system conditions and non-idealities. The backbone encoder and multi-head decoder architecture, combined with continual learning, enables the residual AI model to remain responsive to future changes in the grid. Extensive simulations on the IEEE 68-bus benchmark confirm the effectiveness of this approach, showing improved accuracy in capturing transient behaviors and adaptability across diverse disturbances and predictive architectures. Future work will investigate uncertainty quantification of different adaptation schemes. These results highlight the efficacy of learning-augmented hybrid modeling for enhancing reliability and situational awareness in power system operations.

\section*{Acknowledgment}

The authors acknowledge the assistance of GPT and Claude in numerical implementation and Copilot for editing purposes, with thorough review and oversight by the authors.

\bibliographystyle{ieeetr}
\bibliography{ref}
\end{document}